\documentclass[sigconf]{acmart}
\usepackage{multirow}
\newcommand{\added}[1]{#1}

\AtBeginDocument{%
  \providecommand\BibTeX{{%
    \normalfont B\kern-0.5em{\scshape i\kern-0.25em b}\kern-0.8em\TeX}}}

\copyrightyear{2023}
\acmYear{2023}
\setcopyright{acmlicensed}\acmConference[UIST '23]{The 36th Annual ACM
Symposium on User Interface Software and Technology}{October 29-November
1, 2023}{San Francisco, CA, USA}
\acmBooktitle{The 36th Annual ACM Symposium on User Interface Software
and Technology (UIST '23), October 29-November 1, 2023, San Francisco, CA,
USA}
\acmPrice{15.00}
\acmDOI{10.1145/3586183.3606777}
\acmISBN{979-8-4007-0132-0/23/10}

\begin{document}
\newcommand{\sysc}{PromptPaint}
\newcommand{\inquote}[1]{\textit{``#1''}}
\title{\sysc{}: Steering Text-to-Image Generation Through Paint Medium-like Interactions}


\author{John Joon Young Chung}

\affiliation{%
  \institution{SpaceCraft Inc.}
  \city{Los Angeles}
  \country{USA}}
\email{jjyc@spacecraft.inc}
\authornote{Work done at the University of Michigan.}

\author{Eytan Adar}
\affiliation{%
  \institution{University of Michigan}
  \city{Ann Arbor}
  \country{USA}}
\email{eadar@umich.edu}

% \author{Anonymized}
% \affiliation{%
%   \institution{Anonymized}
%   \streetaddress{Anonymized}
%   \city{Anonymized}
%   \country{Anonymized}}
% \email{Anonymized}

\renewcommand{\shortauthors}{Chung and Adar.}

\begin{abstract}

While diffusion-based text-to-image (T2I) models provide a simple and powerful way to generate images, guiding this generation remains a challenge. For concepts that are difficult to describe through language, users may struggle to create prompts. Moreover, many of these models are built as end-to-end systems, lacking support for iterative shaping of the image. In response, we introduce \sysc{}, which combines T2I generation with interactions that model how we use colored paints. \sysc{} allows users to go beyond language to mix prompts that express challenging concepts. Just as we iteratively tune colors through layered placements of paint on a physical canvas, \sysc{} similarly allows users to apply different prompts to different canvas areas and times of the generative process. Through a set of studies, we characterize different approaches for mixing prompts, design trade-offs, and socio-technical challenges for generative models. With \sysc{} we provide insight into future steerable generative tools.\footnote{The code is available at \url{https://github.com/johnr0/PromptPaint}} 

\end{abstract}





\maketitle
\begin{figure*}
\input{sections/00_teaserfigure.tex}
\end{figure*}
\section{Introduction}
New diffusion-based techniques~\cite{imagen, dalle2, nichol2021glide, stablediffusion} are enabling a wide array of text-to-image (T2I) models. Prompt-driven image creation allows even those without drawing or painting skills to produce high-quality images. Unfortunately, simple text prompts are not always useful for getting what the user imagines in their mind. While the proliferation of cutting-edge tools and demos make new features available (e.g., Midjourney~\cite{Midjourney}, Dream Studio~\cite{dreamstudio}, Gradio demos~\cite{Gradio}, demos in Google Colab Notebooks~\cite{Colaboratory}), guiding them is still challenging.

Artists often create images in a step-by-step procedure: fixing, refining, and improving their ideas as they go. People usually follow specific workflows to produce visual arts, with intermediate decisions between steps~\cite{yan2022flatmagic, dixon2010icandraw, eisner1978children}. A key problem with generative models is that they work largely in an end-to-end fashion: a prompt goes in and an image comes out, with little chance to intervene in between.  For example, in digital comics, artists create the piece in multiple steps: sketching, flatting, shadowing, drawing backgrounds, and adding special effects~\cite{yan2022flatmagic}. Each step allows for refinement and control. AI systems often hide these intermediate steps. Similarly, images generated only with the user's initial prompts would limit what the user can do during artifact production. A second problem is that natural language prompts are not expressive enough for all intents. Just as we would be challenged to describe the art we see, users may find it impossible to describe the art they imagine. This is particularly hard when concepts are ambiguous or don't yet exist (e.g., a style with elements of both Impressionism and Arte Nouveau). The user might not have sufficient natural language descriptions for what they want. Such natural language prompts also lack the ability to specifically control parameters (e.g., how do I get an image with a `flatness' of 60\%?).

To address these challenges, new technical approaches have emerged to enable the gradual editing of visual content. For example, we now see methods to `in-paint' and `out-paint,' adding or revising visual elements on the existing image~\cite{saharia2022palette, avrahami2022blended, nichol2021glide}. 
Users can now also give an initial image to build up on the generation~\cite{avrahami2022blended} or the visual structure~\cite{zhang2023adding}. 
Researchers have also investigated technical approaches to mix prompts~\cite{liu2022compositional, liew2022magicmix} or edit images based on natural language prompts~\cite{hertz2022prompt, kawar2022imagic, couairon2022diffedit, valevski2022unitune}. While the underlying algorithms can help end-users control the images they produce, there is very little consideration for how interactions should be modeled to support the creation experience.

In this work, we explore how users can interact with T2I models to enable the gradual building of artifacts while allowing flexible exploration in the `art space.' To facilitate the steering of T2I models, we take inspiration from how artists interact with paint mediums (e.g., oil paint or watercolor).
The main characteristic of the paint medium we leverage is the flexibility in the use and combination of colors. While we start painting with discrete colors in color tubes, we do not limit ourselves to those tubes but explore colors beyond them by mixing them on the palette. 
Moreover, we apply them on the canvas flexibly, either by overlaying different colors with each other or by using different colors on different canvas areas. 
With \sysc{} we were inspired by this idea and implemented the system to allow users to interact with prompts as they would with colors. \sysc{} turns prompts into flexible materials that can even target verbally indescribable concepts with \textit{prompt mixing} and \textit{directional prompt}. \sysc{} modularizes image generation by allowing users to apply varying prompts to different parts of the canvas (\textit{prompt stencil}) and different parts of the generation process (\textit{prompt \added{intervention}}).

With \sysc{}, we characterized different approaches in their effectiveness for adding new attributes to an existing image. We found that different strategies have different strengths---prompt mixing and directional prompt were effective in adding a new attribute, and prompt \added{intervention} and prompt stencil tend to transform the image while maintaining the visual similarity to the original image. 
We also conducted a user study to identify how users interact with \sysc{}.
From the user study, we found that different ways to steer T2I generation could allow users to generate images that align well with their intentions through iterations. However, we also identified design trade-offs between 1) focused iteration and curation and 2) manual editing and automation. Furthermore, the high complexity and randomness of AI models could result in a misalignment between AI behaviors and user expectations. Lastly, while users had some sense of ownership of the resulting artifacts, their expertise and alignment of the produced artifact with expectations can impact that sense of ownership. From the findings, we discuss insights into adopting paint-medium interactions in designing future versions of generative tools. 

\section{Background and Related Work}

\subsection{Painting From Physical to Digital}

The act of rendering an image by applying paint to canvas is an important form of creative expression~\cite{eisner1978children}. ``Personal causation''~\cite{de1970personal}, or the change in the world by an individual, is an intrinsic satisfaction of painting. Painting enables unique ways of expressing ideas and emotions than other mediums, such as poetry or music. Critically, painting often involves continuous judgments during the process~\cite{eisner1978children}, which can be routinized in a workflow for a specific artifact type~\cite{yan2022flatmagic, dixon2010icandraw}. These characteristics hint at the limitations of existing T2I models. First, not all visual ideas can be described with text. Ideally, the way we craft an image should be closer to the medium itself. Second, the gradual judgments and iterations inherent in the painting are difficult with T2I models. Third, removing the physical act of painting, as T2I models do, reduces the feeling of ``personal causation.'' With \sysc{}, our goal is to address these concerns using painting interactions. The combination of painting interactions with generative approaches supports the balance of direct manipulation with intelligent interfaces~\cite{shneiderman1997direct}.

Researchers have designed many tools for painting and drawing. 
Some tools guided novice users without directly intervening in user drawings~\cite{williford2019drawmyphoto, iarussi2013drawingassistant}.
Others augment by adding corrections to the drawn results~\cite{xie2014portraitsketch, su2014ezsketching, limpaecher2013real, fernquist2011sketch}. 
There are tools to target specific sub-problems in painting (e.g., flexible exploration of colors with color mixing interactions\added{~\cite{shugrina2019colorbuilder, shugrina2017playful}}).
Instead of supporting ``existing drawing/painting practices,'' some systems enabled users to generate novel types of artifacts with computationally enhanced brushes~\cite{benedetti2014painting, jacobs2018extending, sethapakdi2019painting}. 
With AI, systems can now support co-creation, where humans and machines take turns in drawing~\cite{oh2018i, davis2016empirically}.
\sysc{} builds upon these approaches by bringing diffusion-based T2I models closer to interactions with paint mediums. 

\subsection{AI Image Generation}
\label{sec:image_generation}

There have been many approaches to generating images (beyond T2I diffusion models) with neural networks ranging from style transfer algorithms~\cite{gatys2015aneural, dumoulin2016learned, Sheng2018AvatarNet} to generative adversarial networks (GAN)~\cite{Goodfelow2014Generative, choi2018stargan}. 
The most recent approaches include diffusion models that learn to recover images from noisy images~\cite{ho2020denoising}. These generate higher-quality images compared to other approaches, and researchers have devised ways to guide their generation with specific classes~\cite{ho2022classifierfree, dhariwal2021ranzato}.

% contrasive learning --- shared representation of vision and language
In parallel to these techniques, new models include trained representations that combine text and images. CLIP~\cite{radford2021learning}, for example, enables natural language guidance for image generation~\cite{kwon2021clipstyler, crowson2022vqganclip}. Diffusion-based T2I models are some of the most popular due to their flexibility, ability to follow input prompts, and high-quality output~\cite{dalle2, imagen, nichol2021glide}.
However, these models are largely end-to-end in their approach (prompt in, image out). Therefore, imbuing more human intention into the generated results can be challenging. 
Various approaches have tried to tackle this problem, from seeding an initial image to be transformed~\cite{avrahami2022blended, meng2022sdedit} to combining two different prompts to realize them in the image~\cite{liu2022compositional}, generating an image of one prompt while having the overall form of another prompt~\cite{liew2022magicmix}, editing or expanding images with visual masking~\cite{saharia2022palette, avrahami2022blended, nichol2021glide}, editing images with prompts~\cite{hertz2022prompt, kawar2022imagic, couairon2022diffedit, valevski2022unitune, parmar2023zero, brooks2022instructpix2pix}, giving visual structures~\cite{li2023gligen, zhang2023adding, huang2023composer}, and automatically refining prompts~\cite{wang2023reprompt}.
Although these introduced technical approaches to gradually and iteratively shape images, they are largely unconcerned with the interaction model. 
We address this limitation by combining diffusion approaches with novel interaction techniques inspired by physical acts of painting.

\subsection{Interaction with AI Generation}

There are numerous approaches to \textit{steer} AI generations. 
Controls vary from category selection~\cite{krause2020gedi, choi2018stargan} (e.g., happy vs. sad face) to sliders on a fixed continuous semantic scale\added{~\cite{dang2022ganslider, park2020swapping, louie2020noviceai, chiu2020human}} (e.g., melody on a positive-negative scale).
More flexible control of continuous scales includes explorable galleries~\cite{zhang2021method}, user-definable sliders~\cite{chung2023artinter}, or visual sketches~\cite{chung2021talebrush}. Although more flexible, these approaches limit options to a somewhat constrained set.
An alternative to widget-based controls is using examples as inputs~\cite{Sheng2018AvatarNet, park2019arbitrary} (e.g., generating an image similar to the example). 
Although technically flexible in receiving ``any examples,'' steering these models can be challenging as searching for another desirable example can be difficult. 
With advances in language models~\cite{brown2020language} and contrastive learning between text and other mediums~\cite{radford2021learning, wu2021wav2clip, agostinelli2023musiclm}, natural language prompts are used for model steering. 
Prompts can steer generation of texts~\cite{wu2021ai, strobelt2022interactive, brown2020language}, images~\cite{imagen, dalle2, liu2022opal, liu2022compositional}, UI designs~\cite{kim2022stylette}, codes~\cite{chen2021evaluating}, 3D models~\cite{jain2021dreamfields, poole2022dreamfusion}, music~\cite{agostinelli2023musiclm}, and even videos~\cite{makeavideo, phenaki}. 
Prompting has comparative advantages over other approaches, as it does not limit the input the user can make. That is, with the obvious exception that they need to be able to say it. Textual prompts can also be challenging due to: 1) the wide variety of ways to describe something; and 2) the difficulty in describing some concepts due to ambiguity or vagueness (e.g., ``a bit less vivid color'').
Mixing prompts can help overcome these challenges by providing the grounding of a set of `base' textual prompts but with the ability to select the vague semantic spaces between the prompts. 
Some interfaces explored this approach by showing multiple results from prompts with different mixing weights~\cite{automatic1111_2023}. 
\sysc{} adopt paint-medium-like interactions to allow users to visually explore and iterate mixed prompts. 

Finally, we should consider how the generation processes and results are embedded into the human art creation process. 
Generative models should be able to provide intermediate representations~\cite{yan2022flatmagic}, which align with the user workflow and allow easy edits and iterations. 
However, generative models are often designed to produce high-fidelity, final artifacts. 
\added{With GAN models, Endo explored one approach to enable edits, by allowing iteration on high-fidelity image generation with the user's direct manipulation input~\cite{endoPG2022}.}
In diffusion-based T2I model contexts, researchers and practitioners investigated ways to repurpose generation results as a \textit{modularized} unit in the human creation process. As introduced in Section~\ref{sec:image_generation}, editing with seed images, masking, or prompts would be specific examples.
However, not many approaches have looked at how to allow users to intervene during generation. 
We investigate both approaches to 1) repurposing generated \textit{results} as a modularized unit for human artistic creation and 2) allowing user interaction during the generation \textit{process}.  

\section{Interacting with Generative Models Like Paint Medium} 
\label{sec:pp_design}

\added{While the goals of generating artifacts might not directly correspond to those of manual painting, we propose that analogies from painting interactions can facilitate the design of steering interactions for generative models (Figures~\ref{fig:interaction_vector} and \ref{fig:interaction_ingen}).}
We focus on two different aspects. The first is going beyond discrete semantics (e.g., categories, prompts) for specifying generation and flexibly exploring semantics in the vector space.
With the paint-medium analogy, we connect this to color-mixing interactions. 
The second is allowing the gradual generation of artifacts, similar to how we gradually apply colors when we paint. 
In the following, we detail the connections between the steering of generative models and paint-medium interactions. 

\subsection{Mixing Colors: Exploring Vector Spaces}

\begin{figure}[t]
\centering
  \includegraphics[width=0.478\textwidth]{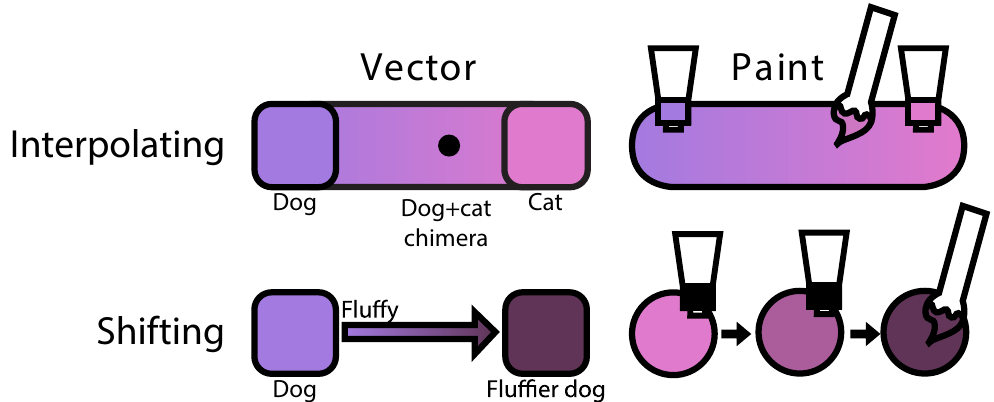}
  \caption{
  Mapping vector space exploration to paint color mixing. Discrete semantics (e.g., categories, prompts, or examples) are represented as a rounded square in Vector. They can map to discrete color tubes in Paint. Using the analogy, the user can explore semantics between discrete ones in a way similar to how they would explore colors by mixing.
  }
  \Description{On the top, there is a part of the figure explaining interpolation. For vectors, two different concepts, dog and cat, are connected with a gradient line, and one of the middle is selected with a black dot, which is labeled as Dog+cat chimera. On the right of that, there is a analogical figure on paint, which has two colors mixed and a brush picking a color within the mixed color space. On the bottom, there is a part explaining shifting. For vectors, is is showing the prompt of concept, dog, and there is an shifting vector arrow of fluffy, which leads to the concept of fluffier dog. For the analogical figure on paint, a black color is being added to a color, which gradually turns into darker colors, with the final color picked by a brush. }
  \label{fig:interaction_vector}
\end{figure}

Discrete input modalities of categories, examples, and prompts specify the semantics of generation while being easily comprehended by users. Generative models, such as language models~\cite{krause2020gedi, chung2021talebrush}, style transfer algorithms~\cite{Sheng2018AvatarNet}, GANs~\cite{alaluf2021hyperstyle, karras2019astylebased}, or diffusion models~\cite{dalle2, imagen}, first transform these inputs into vector representations. 
As a vector is a continuous representation, describing representations \textit{between} discrete semantics would be difficult with only discrete interfaces (e.g., \added{varying degress of `chimeras'} of \textit{cat} and a \textit{dog}). 
However, there are cases where users want to work with such semantics. For example, in some situations that are difficult to verbally describe, users might want to use vector representation spaces. Such a need would also arise when the user wants to do fine-grained control of an attribute, like adjusting the roughness of image textures or the fluffiness of a dog. 
\added{Moreover, exploration of intermediate semantics would facilitate realizing eclecticism, where the artist tries to mix different styles together~\cite{hume2010art}.} 
Previous work has shown that such manipulation is doable by interpolating discrete semantics\added{~\cite{chung2021talebrush, radford2016unsupervised, kim2022mixplorer}} or \added{shifting semantics} with directional vectors about concepts~\cite{park2020swapping, schwettmann2021toward}. However, these approaches have not generally offered ways of turning vector manipulations into accessible interactions, specifically when the user can flexibly specify different discrete semantics (e.g., prompts).

To explore vector representations, we introduce the idea of interacting with discrete semantics in a way similar to how we \textit{mix} physical paint colors. When paints are used, they come into our hands in color tubes, each having one discrete color. However, when applying them to the canvas, we do not limit ourselves to those discrete choices. Rather, we create new colors by mixing on a palette. 
Using this idea, we introduce the interaction of mixing different discrete semantics in a \textit{semantic palette}.
Analogically, each discrete semantic of categories, examples, or prompts would map to a discrete color for the paint mediums (Figure~\ref{fig:interaction_vector}). 

The first specific approach to mixing discrete semantics is to \textit{interpolate} them, by mixing two or more discrete semantics on the semantic palette and exploring the space between them (top row of Figure~\ref{fig:interaction_vector}). 
For example, to render an image of a chimera of a cat and a dog with T2I models, the user would interpolate the semantics of a cat and a dog. 
This would be similar to spreading colors to mix them and using intermediate gradients of those colors. 

The second approach is \textit{\added{shifting}}---adding a directional semantic for fine-grained control (bottom row of Figure~\ref{fig:interaction_vector}). 
With a paint medium, this would be like adding a small amount of a different color to change the characteristics of the used color (e.g., making the green color darker by adding a bit of black pigment). 
In our case, the user can render an image of a dog with a certain level of fluffiness by adding the semantics (analogically, ``pigments'') of fluffiness to the semantics of a dog.
Note that these interactions could be adapted to those generative techniques that can turn discrete `user-facing' concepts into the vector space and then perform generation with the vector representations. 

% mixing on the visual spatial space can be more intuitive
    % as their relationships would be represented as whether they are close or far from the original discrete semantics
    % manipulation can be simpler - 3 prompt example
        % more prompts, can be simpler based on the geometry

\begin{figure}[t]
\centering
  \includegraphics[width=0.478\textwidth]{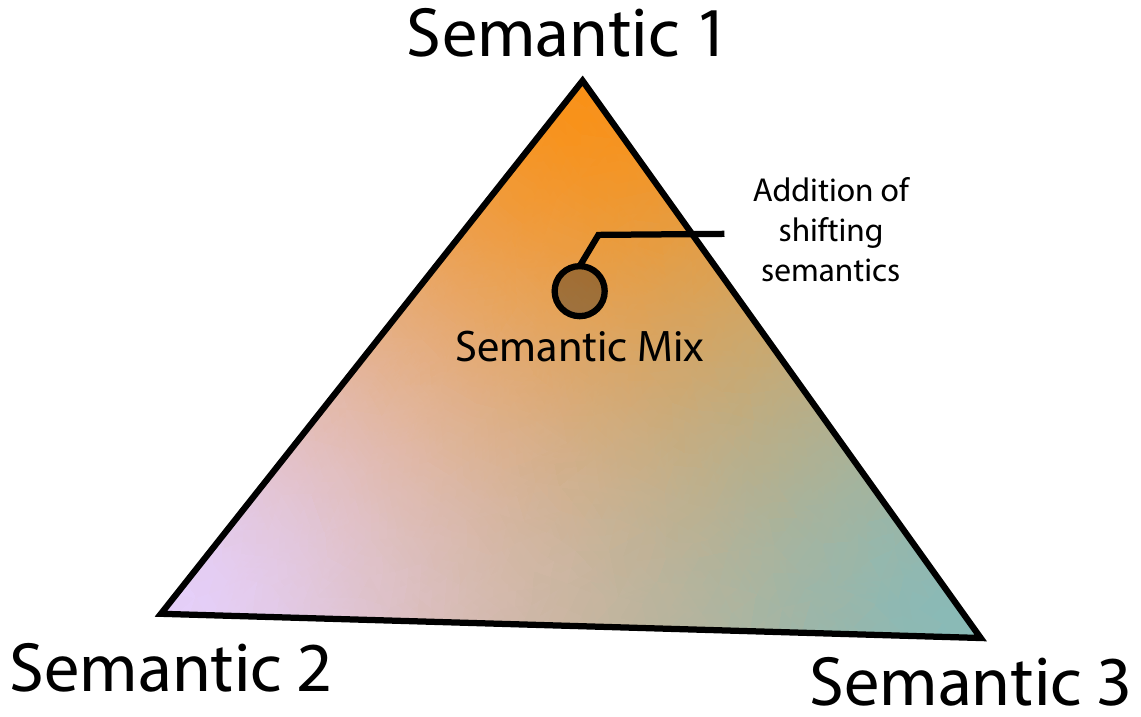}
  \caption{Using palette interaction for semantic mix has the benefit in that the interpolation and the \added{shifting} can be represented in the same interface.}
  \Description{There is a triangle with each corner saying "Semantic 1," "Semantic 2," and "Semantic 3." Three corners are filled with different colors and space between those corners are filled with interpolated gradient color. One of the gradient space is selected with the circle of "Semantic Mix." The circle is annotated as "Addition of shifting semantics."}
  \label{fig:palette_demonstration}
\end{figure}

Naturally, there can be other interactions to mix discrete semantics. For example, we can mix prompts with sliders, each representing the weight of each prompt. Compared to such interactions, palette interactions can represent the mix of semantics with two visual signals: positions and colors on the palette. With palettes, specifically, both interpolation and \added{shifting} can be shown in a single interface. As in Figure~\ref{fig:palette_demonstration}, the palette interface can represent the interpolation with a point in the mixed-color gradient and show the \added{shifting} by adding the color to the selected interpolated point.
On the other hand, conventional sliders may become more complex as the weights for interpolation and \added{shifting} would need to be represented in separate sliders. 

\subsection{Colors onto Canvas: Gradual Generation}

\begin{figure}[t]
\centering
  \includegraphics[width=0.478\textwidth]{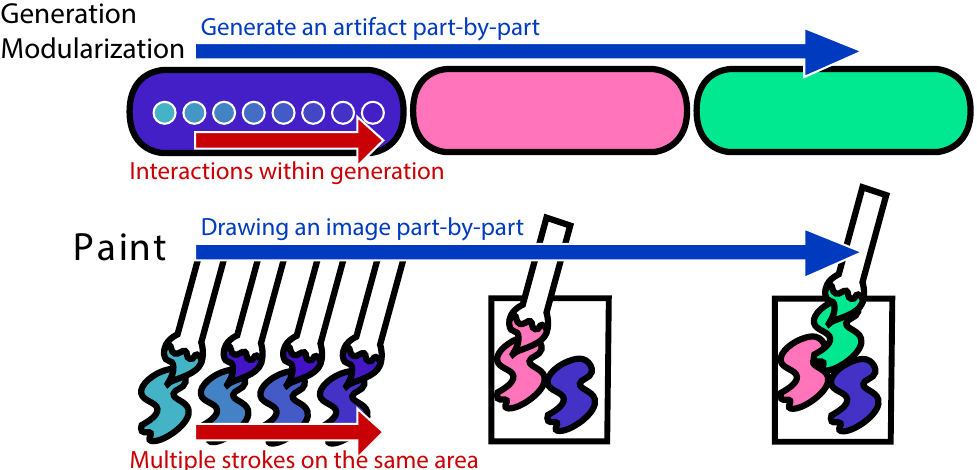}
  \caption{Mapping generation modularization to gradual painting of an artifact. Within-generation interventions would correspond to multiple strokes applied on the same canvas area, and generating the artifact part-by-part would map to drawing the image part-by-part.}
  \Description{The top row is titled as "Generation Modularizaton," and there are big boxes, which stand for generation of a part of the artifact. Inside the first box,there is a series of circles, with gradually changing colors. They indicate interactions during the generation process. These circles are aligned long a red arrow annotated as "Interactions within generation." This box is aligned with other boxes along a blue arrow annotated as "Generate an artifact part-by-part." The below row is titled as "Paint," which is a analogy of the top row to painting. The first part of the figure is showing a brush painting multiple strokes on the same area with different colors. The progress is shown with a red arrow annotated as "Multiple strokes on the same area." This figure is continued with the canvas filling with brush strokes on other areas. The progress is shown with a blue arrow annotated as "Drawing an image part-by-part."}
  \label{fig:interaction_ingen}
\end{figure}

Generation models do not often allow user interventions during the generative process. Thus, the experience of using generative models can be far from ``creation,'' where the painter gradually shapes the artifact, making decisions as they go. Instead, we suggest that interventions could be applied by the end-user \textit{during} the generative process.
Again, we take the analogy of painting, focusing on how we apply paint to the canvas. 
When we apply colors to the canvas, we do not use the same paint for the whole area. 
Instead, we gradually build the artifact with multiple paint strokes and overlapping layers. We propose that modularizing the generation model temporally and spatially would allow for interactive changes to steer the model.

We consider two forms of modularization. 
The first allows interactions (temporally) within the generation process (red arrows in Figure~\ref{fig:interaction_ingen}). One example interaction can be changing the guiding prompt \textit{during} generation. In our paint metaphor, this would be similar to overlaying different paint strokes on the same area to decide the final rendition. Not all models support this kind of intervention, though the diffusion-based T2I model does. As we will demonstrate, in diffusion-based T2I models, prompts that are used in the earlier stage can decide the overall form of the image while those in the later part decide the details. For example, brushing with a ``banana on the ground'' prompt-as-color first and then switching to ``a futuristic car'' color would result in a futuristic car in the shape of a banana. 

The second form of modularization is the spatially partial generation of content (blue arrows in Figure~\ref{fig:interaction_ingen}). Analogically, this would be equivalent to how people draw an image part by part. 
Again, not all models are capable of this kind of focused generation. However, in-painting and out-painting in diffusion-based T2I model can support this functionality~\cite{saharia2022palette, avrahami2022blended, nichol2021glide}. For example, with T2I models, the user can first generate the overall background and specific objects later. In the language of prompts-as-color, a brush could be loaded with an `ocean' color and applied to the background to be followed by the targeted application of the `boat'-color to certain areas.
\section{\sysc{}: Interface}

Using the interactions described in Section~\ref{sec:pp_design}, we built \sysc{}, an image creation tool powered by a diffusion-based T2I model (Figure~\ref{fig:pp_teaser}). \sysc{} supports the flexible steering of the generative model with 1) exploration and fine-grained control of prompt space with \textit{prompt mixing} and \textit{directional prompts}, and 2) the gradual building of images with \textit{prompt \added{intervention}} and \textit{prompt stencils}. 

\subsection{Canvas and Basic Editing Functions}
\sysc{} presents a canvas where the user can create raster images (Figure~\ref{fig:pp_teaser}h) with basic image editing functions. This includes moving/rotating images inside the canvas, brushing, erasing, and lassoing (from left to right of Figure~\ref{fig:pp_teaser}a, except the right two). Furthermore, the user can add layers, change their ordering, hide, or even delete them (Figure~\ref{fig:pp_teaser}h). 

\subsection{T2I Generation Functions}
Through diffusion-based T2I functions, the user can generate images on the canvas. The user first specifies the prompts to guide the generation (\textit{prompt mixing} and \textit{directional prompt}). The user can then start the generation by specifying the area to which generation results should be applied (\textit{prompt stencil}). During generation, the user can also change the guiding prompts (\textit{prompt \added{intervention}}). 

\subsubsection{Prompt List}
The user can add prompts in the Prompt List (Figure~\ref{fig:pp_teaser}b). They can add a new prompt with the \textsc{+} button. Each added prompt has its own color (editable through a color picker) and editable prompt text. The user can delete the prompt with the \textsc{X} button.

\subsubsection{Prompt Mixing}

\begin{figure}[t]
\centering
  \includegraphics[width=0.478\textwidth]{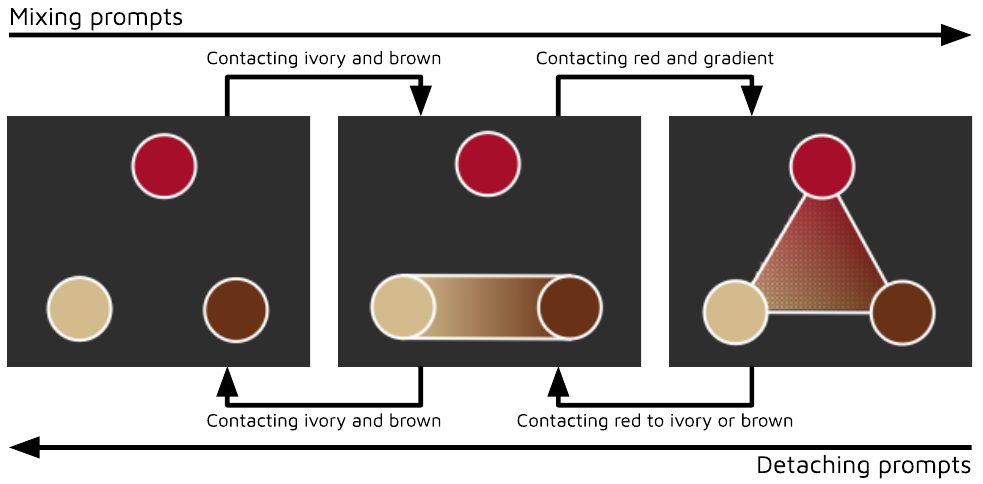}
  \caption{Interactions to mix/detach prompts in the Prompt Palette.}
  \Description{There are figures for three states in colors ivory, brown and red. These states are aligned horizontally, with the right direction indicating "mixing prompts" while the left direction indicating "detaching prompts." The first state has three circles. The second state has three circles, while two connected with a gradient color. The third state also has three circles, with all three connected with a triangle with the mixed gradient colors. From the first state to the second state, there is an arrow saying "contacting ivory and brown." From the second to third state, there is an arrow "contacting red and gradient." From the third to second, there is an arrow "contacting red to ivory or brown." From the second to the first, there is an arrow "contacting ivory and brown." }
  \label{fig:pp_prompt_mixing}
\end{figure}

\begin{figure}[t]
\centering
  \includegraphics[width=0.478\textwidth]{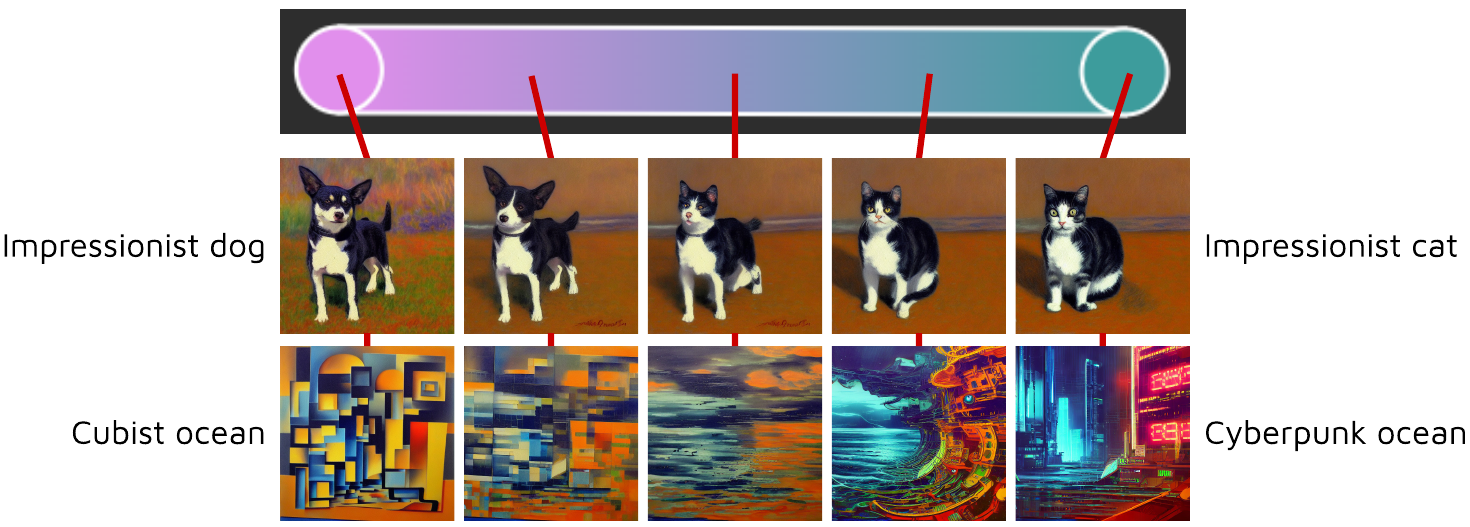}
  \caption{Example results of prompt mixing.}
  \Description{In the upper part of the figure, there is a colored gradient fro pink to blue. Below that, on the left side, there is an image of a dog with a caption "impressionist dog." On the right side, there is an image of a cat with a caption "impressionist cat." In between these two, there are images that are gradually turning from dog to cat, from the left to right. Each image is connecte to the upper gradient, showing the change in the mix of prompts. Below this row, on left, there is an image of "Cubist ocean," and on the right, there is an image of "Cyberpunk ocean." Between them, there are image with the mixed style from cubist ocean to cyberpunk oceans, from left to right. There images are also connected to the color gradient. }
  \label{fig:pp_prompt_mixing_demonstration}
\end{figure}

\sysc{} renders each prompt as a circle in a palette area (Figure~\ref{fig:pp_teaser}c). 
The user can move and organize these prompts by dragging them (just as they could decide where to place their paints on a regular palette).
The user can `blend' the prompts though \textit{prompt mixing} to explore the vector space between specified ``discrete'' prompts\added{~\cite{kim2022mixplorer}}. To do this, the user can directly manipulate prompt color circles with an interaction similar to how we mix paint mediums~\cite{shugrina2019colorbuilder}. As in Figure~\ref{fig:pp_prompt_mixing}, the user can touch one of the prompts on another to mix two prompts. If the user wants to add a third prompt to the mix, they can touch the already mixed gradient with the third one. 
While the current version of \sysc{} allows mixing a maximum of three prompts, future versions could allow mixing more. 
If the user wants to detach a prompt from the mix, they can touch the prompt with another prompt within the mix. 
For a generation input, the user can select one of the prompts or a point from a mixed gradient of prompts. The selected will be rendered as a circle (Figure~\ref{fig:pp_teaser}c-2). The Prompt List will highlight the prompts mixed in the selection in green (Figure~\ref{fig:pp_teaser}b). 
\sysc{} would interpolate these prompts to guide the diffusion model's generation. Figure~\ref{fig:pp_prompt_mixing_demonstration} shows examples of mixing two different prompts with prompt mixing. 

\subsubsection{Directional Prompt}

\begin{figure}[t]
\centering
  \includegraphics[width=0.478\textwidth]{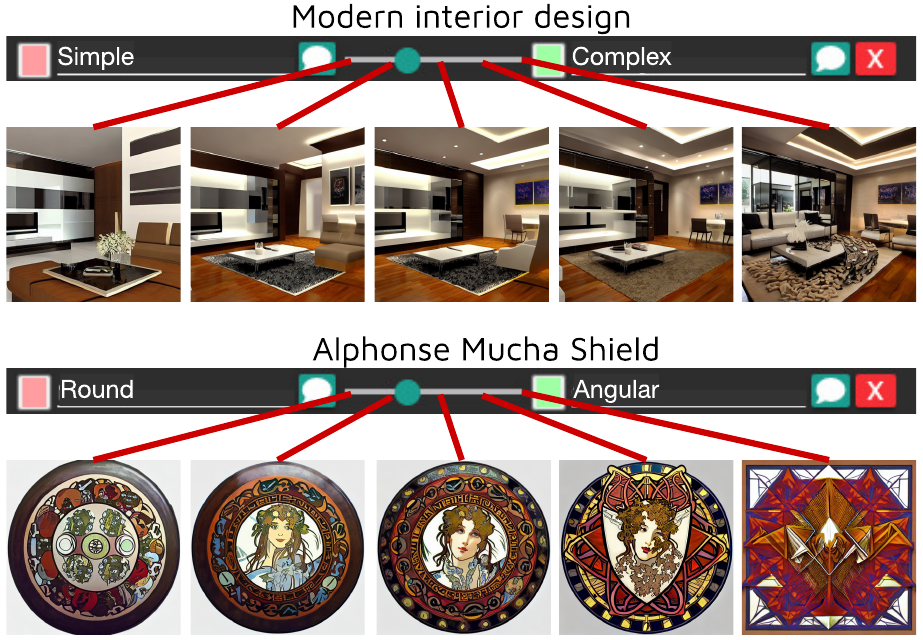}
  \caption{Example results of directional prompts. The center is the result without a directional prompt and the left and the right are results of applying directional prompts. The rightmost and leftmost results applied the full vector difference between the two end prompts.}
  \Description{The first part of the image is titled as "Modern Interior Design." Below the title, there is a slider, with two different ends of "Simple" and "Complex." Below that, there are images that correspond to different part of the slider, that shows interior designs in varying complexities. The second part of the image is titled as "Alphonse Mucha Shield." Below the title, there is also a slider, with two different ends of "Round" and "Angular." Below that, there are images correspond to different part of the slider, that shows shield design with varying levels of roundness.}
  \label{fig:pp_directional_prompt_demonstration}
\end{figure}

Directional prompts allow users to \added{shift} the prompts \added{by introducing} additional attributes\added{~\cite{park2020swapping, schwettmann2021toward}}, with interactions similar to adding other colors (Figure~\ref{fig:pp_teaser}d). 
The user can add a new directional prompt with the \textsc{+} button. With it, they can set two ends with prompts and decide the direction of the attribute to add. Each end has a unique, user-definable, color. 
After setting the two ends, the user can toggle the slider to set the intensity of the attribute to add. In Figure~\ref{fig:pp_teaser}d, for example, the user can add a slight amount of the ``matte'' attribute by moving the slider closer to ``matte.'' As the user moves the slider to one end in the directional prompt, the background color of the Prompt Palette gradually changes to the unique color of the end. With multiple directional prompts, this color changes to a mix of colors from the ends, with weights according to the slider values. Figure~\ref{fig:pp_directional_prompt_demonstration} shows examples of using directional prompts.

\subsubsection{Prompt Stencil}

\begin{figure}[t]
\centering
  \includegraphics[width=0.478\textwidth]{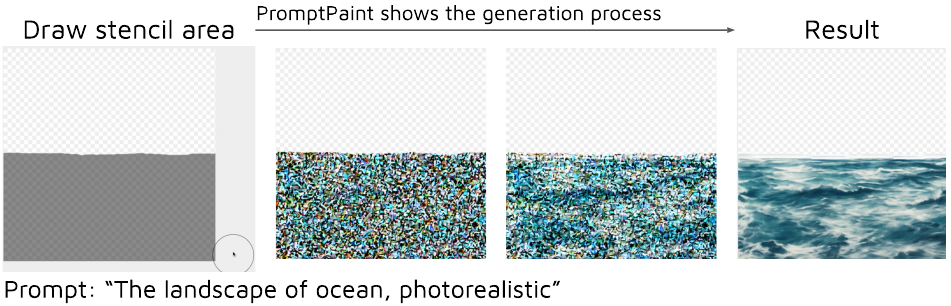}
  \caption{With a prompt stencil, the user can specify the area of generation with brushing (dark grey). When the user completes brushing, the tool starts generating a part of the image while showing the process to the user.}
  \Description{On the left, there is an image of a canvas with specified area, and the image is titled "Draw stencil area." In the bottom of the image, there is a prompt, "the landscape of ocean, photorealistic." On the right, there are images that show graudal change of the image with the diffusion process. Images are becoming more and more concrete from a very noisy image. There is an arrow that aligns these images, and the arrow is annotated "PromptPaint shows the generation process." The far right image is a realistic ocean scene.}
  \label{fig:pp_stencil_demonstration}
\end{figure}

After setting the prompt to use, the user can gradually build the visual images with a prompt stencil. 
The user can specify the area of the generation with brushing\added{~\cite{saharia2022palette, avrahami2022blended, nichol2021glide}} while \textsc{Draw} is selected (Figure~\ref{fig:pp_teaser}e). As the user completes brushing, \sysc{} starts to generate an image, with intermediate generation results shown to the user in real-time (Figure~\ref{fig:pp_stencil_demonstration}). The progress bar in Figure~\ref{fig:pp_teaser}g shows how much of the generation is done. 
The user can repeat this process to fill in other canvas areas. 
Note that the user can adjust the intensity of the guidance and the number of intermediate steps with \textsc{guide Scale} and \textsc{steps} in Figure~\ref{fig:pp_teaser}e.

\begin{figure}[t]
\centering
  \includegraphics[width=0.478\textwidth]{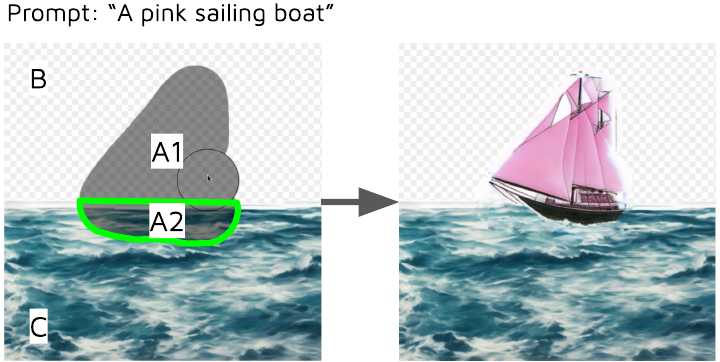}
  \caption{When applying prompt stencil upon the existing image, for the area where the stencil is overlapping with the existing image (A2), \sysc{} generates a new image that is similar to the existing image based on the overcoat value (higher, less similar).}
  \Description{On the left, there is an image that has the bottom half filled with an ocean image. And there is also a stencil in the shape of a boat. The part of the stencil that is not overlapped with the ocean (or stenciled empty part) is annotated "A1." The part that is overlapping with the ocean is annotated "A2." The empty part that is not stenciled is annotated "B." The ocean part that is not stenciled is annotated as "C." On the right, there is a pink boat rendered on the ocean part.}
  \label{fig:pp_multiple_stencil}
\end{figure}

\begin{figure}[t]
\centering
  \includegraphics[width=0.478\textwidth]{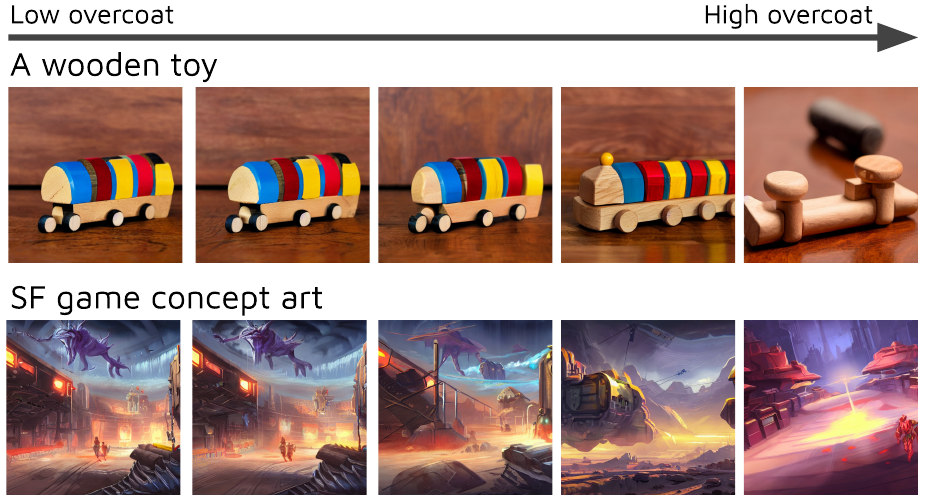}
  \caption{Example results of overcoating generation. Images are generated again from the far left image with varying overcoat values (more right, higher overcoat values). Images generated with higher overcoat values tend to be less similar to the original image.}
  \Description{On the top, there is an arrow starting from "low overcoat" to "high overcoat." Below that, there is one set of images of wooden toys and another set of images of SF game concept art. From left to right, the image is gradually changing from one version of wooden toy/SF game concept art to another version.}
  \label{fig:pp_overcoat_demonstration}
\end{figure}

When the canvas already has images on the layer, \sysc{} considers those existing content to generate new content. 
For example, as in Figure~\ref{fig:pp_multiple_stencil}A2, when the user's new stencil overlaps with the existing images, based on the \textsc{overcoat} value (Figure~\ref{fig:pp_teaser}e), \sysc{} tries to generate a new image that is similar to the already generated images (the higher the overcoat value, the less similar). Figure~\ref{fig:pp_overcoat_demonstration} shows the impact of varying overcoat values when generating an image again with the same prompt. We can also use overcoating to change an image by using different prompts from those used to generate the existing image.

\subsubsection{Prompt \added{Intervention}}

\begin{figure}[t]
\centering
  \includegraphics[width=0.478\textwidth]{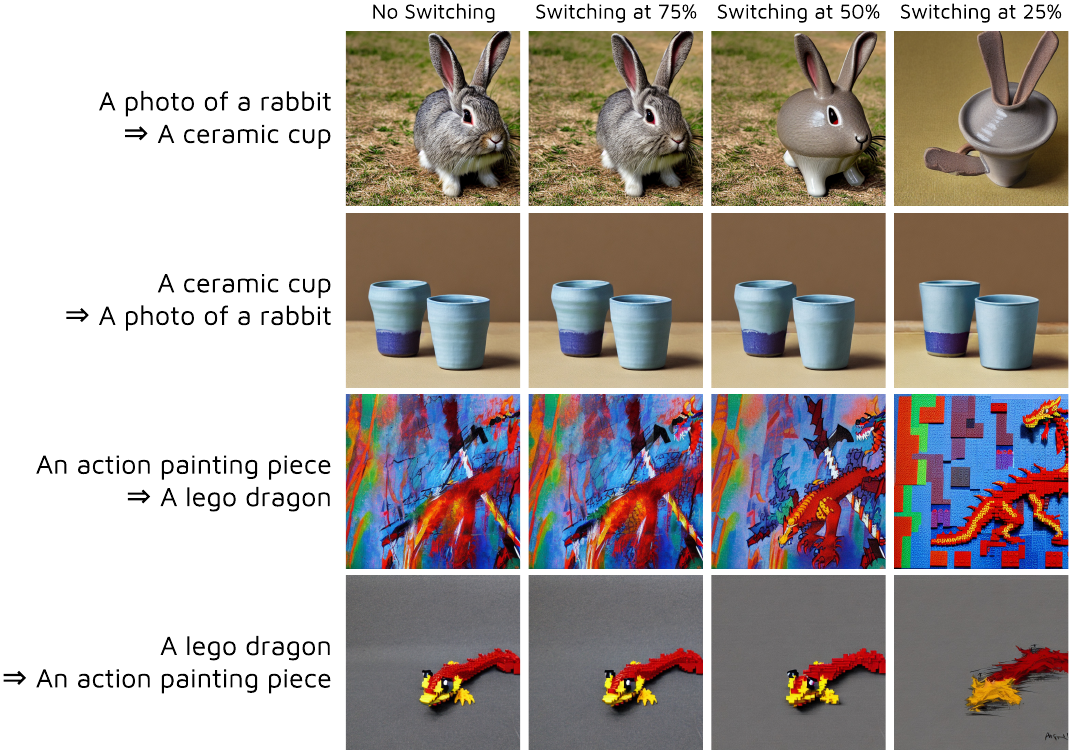}
  \caption{Example results of prompt \added{intervention}. Except for the first column which did not change the prompt during the generation, each column switched the guiding prompt at a different point of the generation process.}
  \Description{The image has four rows and four columns, where each row stands for a case of prompt intervention and each column stands for different levels of prompt intervention. The first row shows changing the prompt from "A photo of a rabbit" to "A ceramic cup," and the image is gradually changing from a realistic rabbit to a rabbit-shaped cup. The second row is changing the prompt from "A ceramic cup" to "A photo of a rabbit," and the images are not drastically changing for this case. The third row is from "An action painting piece" to "A lego dragon," and the image is gradually changing from an action painting piece to similarly shaped lego dragon. The last row is from "A lego dragon" to "An action painting piece," where the image is gradually changing from a lego dragon to similarly shaped action painting piece.}
  \label{fig:pp_switching_demonstration}
\end{figure}

With \textit{prompt \added{intervention}}, \sysc{} allows interactions \textit{during} the generation process. The user can change either 1) the selection of the prompt in the Prompt Palette or 2) the slider values for directional prompts. 
With the change of prompts, as in Figure~\ref{fig:pp_switching_demonstration}, prompts used in the earlier stage of the generation tend to decide the overall form and color, while those used in the later stage decide details. 
Hence, this technique can maintain the visual form of the generation with iterations. However, as in the second row of Figure~\ref{fig:pp_switching_demonstration}, sometimes the generation result does not change much even with early prompt \added{intervention}.

As the user can change the prompts during generation, \sysc{} allows for control over the generation process. 
First, \sysc{} visualizes the prompts used in previous generations as paths in the palette interface as in Figure~\ref{fig:pp_teaser}c-1. Note that the used directional prompts decide the colors of the dots. This visualization of past paths helps users understand what they have tried and eases iteration on different combinations of prompts. 
Furthermore, they can stop and restart the generation (the button in Figure~\ref{fig:pp_teaser}g changes to either \textsc{stop} or \textsc{start}). When the user has stopped generation, they can roll back the generation to a specific step, either by selecting the past point in the progress bar (Figure~\ref{fig:pp_teaser}g) or undoing with \textsc{ctrl} and \textsc{z} keys. If the user wants to switch to past versions of generations, they can click one of the dots on the past paths. \sysc{} highlights the dot for the current generation step with a green border. When restarting the generation, the user can also set the number of steps processed in a single ``round'' of generation, with \textsc{single stroke} in Figure~\ref{fig:pp_teaser}e. 
\section{\sysc{}: Technical Details}
\label{sec:technical}
We describe the technical details of \sysc{}. First, we give an overview of diffusion-based T2I generation models. Then, we explain the technical approaches for each function and the implementation details. 

\subsection{Background: Diffusion-based T2I}

\begin{figure}[t]
\centering
  \includegraphics[width=0.478\textwidth]{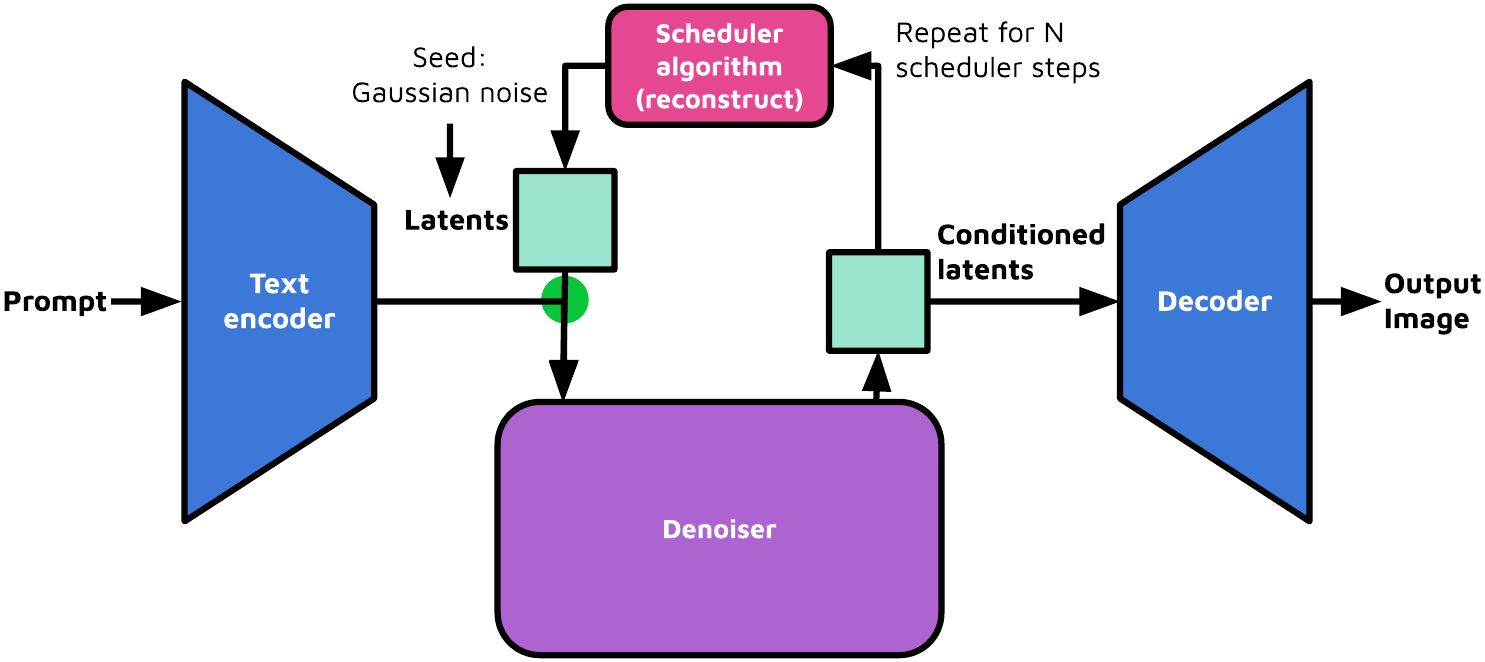}
  \caption{The pipeline of diffusion-based text-to-image models. It shows a specific version, which processes the diffusion process in latent representation. The technical manipulation of \sysc{} happens in the green dot, either by manipulating vector-embedded prompts or intermediate latent. }
  \Description{The left part of the image has "prompt" as an input, connected to a "text encoder" block. This block is connected to "denoiser," which also gets an input from "latent" block. This latent block is connected to "Seed: Gaussian noise," which is used in the initialization. The denoiser output "conditioned latents," which connects to "scheduler algorithm (reconstruct)." The connect arrow is annotated "Repeat for N scheduler steps." The scheduler algorithm then outputs the latents, which again becomes the input to the denoiser. Conditioned latents is also connected to "decoder", which outputs "output image." The input part of denoiser is marked with green dot, indicating PromptPaint's interactions happens in this part.}
  \label{fig:pp_diffusion_diagram}
\end{figure}

% background of diffusion-based model
%We explain a high-level concept of how diffusion-based T2I models work during inference (i.e., during generation). 
While there are many variants of diffusion-based T2I generation models~\cite{dalle2, imagen}, we focus our explanation on the latent diffusion models~\cite{stablediffusion} we used.
Latent diffusion models~\cite{dalle2, imagen, stablediffusion} can largely be characterized by (Figure~\ref{fig:pp_diffusion_diagram}): 1) a text encoder that converts text prompts into vectors used to guide diffusion, 2) a denoiser and scheduler, which gradually process image generation by reducing noise in the latent vectors of the image, and 3) a decoder, which turns latent vectors into a higher resolution output image. 
Note that the decoder is not the universal feature of diffusion-based T2I models, but helps with computational efficiency by processing images in lower-dimension latent representations. 
Often the geometry of the latent representation corresponds to the output image, whereas the output image tends to have higher dimensions. 
When the image on the canvas needs to be used for diffusion (e.g., overcoating), it requires an encoder that encodes the image into a latent representation.
All technical manipulations of \sysc{} occur in the green dot of Figure~\ref{fig:pp_diffusion_diagram}, either by manipulating vector-encoded text prompts (prompt mixing, directional prompts, prompt \added{intervention}) or latent representations from the denoiser and scheduler (prompt stencil). 

\subsection{Prompt Vector Manipulations}
\textit{Prompt mixing} and \textit{directional prompt} manipulate the prompts in the vector space embedded by the text encoder. Prompt mixing interpolates different prompt vectors with weights~\cite{dalle2} from the user's Prompt Palette (with proximity to each prompt):
\begin{equation}
    v_{p_m} = \sum^N w_{p_i} v_{p_i}
\end{equation}
$v_{p_m}$, $w_{p_i}$, and $v_{p_i}$ represent the interpolated vector, the weight of each prompt, and the embedded vector of each prompt, respectively.
The directional prompt first calculates the directional vector between two different prompts: 
\begin{equation}
    v_{d_j} = v_{d_j1} - v_{d_j2}    
\end{equation}
where $v_{d_j}$ indicates the directional vector and $v_{d_j1}$ and $v_{d_j2}$ stand for the embedding of the prompts at both ends. Then, with weights on the slider interfaces ($w_{d_j}$), \sysc{} calculates the final input of the prompt vector:
\begin{equation}
    v_{f} = v_{p_m} + \sum^M w_{d_j} v_{d_j}
\end{equation}
Then, \sysc{} would use $v_f$ as input to the denoiser model to guide the denoising process. 
Users can change $v_f$ during generation, which is how \textit{prompt \added{intervention}} is technically done. 

\subsection{Latent Representation Manipulation}

\textit{Prompt stencil} and overcoated generation require manipulation in latent representations before the denoiser process. Specifically, \sysc{} manipulates latent representations considering different areas with or without stencils and existing image content.
This approach is similar to image-to-image diffusion in previous work~\cite{avrahami2022blended}. 
For the area that is stenciled but does not have any image content (Figure~\ref{fig:pp_overcoat_demonstration}A1), no manipulation is performed. However, for the stenciled area with image content (Figure~\ref{fig:pp_overcoat_demonstration}A2), latent representation is manipulated as follows:
\begin{equation}
    l_{m, k}(x, y)  =\begin{cases}
    \nu(l_I(x, y), k), & \text{if $k<(1-o/100)*K$}.  \\
    l_k(x, y), & \text{if $k>=(1-o/100)*K$}.  
    \end{cases}
\end{equation}
As in the above equation, when the diffusion step ($k$) is smaller than the threshold $(1-o/100)*K$ ($o$ is the overcoat ratio and $K$ is the number of all diffusion steps), $l_{m, k}(x, y)$ (the latent representation after manipulation at the position of $(x, y)$ and the step of $k$) is replaced by $\nu(l_I(x, y), k)$ (where $l_I$ is the latent representation of the existing image and $\nu(l, k)$ is a function that adds noise to the latent representation to the amount adequate to step $k$). Otherwise, the latent representation would not be manipulated and \sysc{} would use the reconstructed latent representation from the scheduler algorithm. 
For unstenciled areas, the area of Figure~\ref{fig:pp_multiple_stencil}B is considered to have a white background. Then, the area of Figure~\ref{fig:pp_multiple_stencil}B and C would be replaced as follows:
\begin{equation}
    l_{m, k}(x, y)  = \nu(l_I(x, y), k)
\end{equation}
Therefore, these unstenciled areas are replaced by the latent representation of existing images with noise added for each step. 
\added{Note that we adopted this approach instead of inpainting models~\cite{suvorov2021resolution} to simulate ``overcoating'' effects where an already-filled area needs to maintain visual similarity with the additional generation on the area.}

\subsection{Implementation}

We implemented \sysc{} as a web app using HTML, CSS, JavaScript, and React. For deploying a diffusion-based T2I model, we built a WebSocket-based Flask server, as \sysc{} shows intermediate generation results in real-time. For the T2I model, we used Stable Diffusion~\cite{stablediffusion}, which uses UNet for the denoiser and the variational autoencoder for the decoder and encoder. For the UNet and the variational autoencoder models, we used 
\texttt{
    \hyphenchar\font=`\- % enable hyphenation
    \hyphenpenalty=10000 % disable it again
    \exhyphenpenalty=-50 % enable (encourage) it at explicit hyphens
runwayml/stable-diffusion-v1-5}\footnote{\url{https://huggingface.co/runwayml/stable-diffusion-v1-5}}. For the text encoder, we used the 
\texttt{openai/clip-vit-large-patch14
\hyphenchar\font=-1} checkpoint\footnote{\url{https://huggingface.co/openai/clip-vit-large-patch14}} of CLIP~\cite{radford2021learning}. For the scheduler, we used the DDIM scheduler~\cite{song2021denoising} with a beta start of 8.5e-4, a beta end of 1.2e-2, and a scaled linear beta schedule. 
\section{Characterization Study}
\label{sec:characterization}
Through a crowdsourced study, we characterize \sysc{} functions in terms of how they allow users to iterate on the already generated images by adding another attribute. 

\subsection{Conditions}
We considered 1) \textbf{prompt mixing}, 2) \textbf{directional prompt}, 3) \textbf{prompt stencil}, 4) \textbf{prompt \added{intervention}}, and 5) \textbf{prompt concatenation}. While the prompt stencil's fundamental purpose is not to add another attribute to existing images, we can repurpose this to ``overcoat'' other visual elements on existing images. The last condition, prompt concatenation, is textually adding another attribute to the existing prompt (e.g., ''mix of impressionism and cubist''). 

\subsection{Dataset}

\begin{table}[t]
\caption{Attributes used in characterization study.}
\begin{tabular}{p{0.07\textwidth}p{0.368\textwidth}}
Type                       & Attributes                                                                                                  \\ \noalign{\global\arrayrulewidth=0.3mm}
\hline
\noalign{\global\arrayrulewidth=0.15mm}
Objects                    & tree, river, man, woman, dog, cat, love, hate                                                           \\ \hline
Styles                     & cubist, surrealism, action painting, high renaissance, impressionism, cyberpunk, unreal engine, VSCO        \\ \hline
Specific visual attributes & vivid color, subtle color, rough texture, smooth texture, fine line, thick line, curvy shape, angular shape
\end{tabular}
\Description{The first row headers for columns, "Type" and "Attributes." The first type is "objects," with attributes of "tree, river, man, woman, dog, cat, love, and hate." The second type is "styles," with attributes of "cubist, surrealism, action painting, high renaissance, impressionism, cyberpunk, unreal engine, and VSCO." The third type is "specific visual attributes," with attributes of "vivid color, subtle color, rough texture, smooth texture, fine line, thick line, curvy shape, and angular shape."}
\label{tab:attributes}
\end{table}

To characterize each condition, we generated an experimental dataset. When adding a new attribute, we considered three attribute types: 1) \textbf{objects}, 2) \textbf{styles}, and 3) \textbf{specific visual attributes}. 
We considered \textbf{objects} and \textbf{styles}, as they are often used as the most basic attributes in T2I generation~\cite{liu2022design}.
We additionally considered specific visual attributes such as colors, textures, lines, and shapes, as they are widely used to describe visual attributes in the practice of visual arts~\cite{elementsofart}. 
Table~\ref{tab:attributes} shows the descriptors we used for each type. We used a subset of object and style descriptors from Liu and Chilton~\cite{liu2022design}. We sample specific visual descriptors from art learning materials~\cite{elementsofart}. 

With these attributes, we systematically generated pairs of images: an image generated with the initial prompt and an iterated version that added another attribute. 
First, we chose a ``target attribute set'' from objects, styles, and specific visual attributes, which would be the type of attribute added in the iterated image. 
Then, we sampled two attributes from the target attribute set, the ``original attribute,'' to be included in the initial prompt, and the ``additional attribute,'' to be added in the iterated image. 
Note that some pairs of attributes can be semantically more relevant to each other than other pairs (e.g., man and woman). Below, we show how the generation can be different between more and less relevant pairs.
To generate images, we needed attributes other than the target attribute and the seed to initialize the noisy latent representation. For example, when we use objects as target attributes, we would need to have style attributes in the prompts. On the other hand, when we pick specific visual attributes as target attributes, both style and object attributes would be required. 
For a pair of target attributes, we randomly sampled two sets of other attributes and seeds.

For each set of attributes (original, additional, and non-target) and seed, we generated images with varying weights and conditions. For \textbf{prompt mixing}, with varying weights, we interpolated vector embeddings of two prompts (one with an original attribute and non-target attributes and the other with an additional attribute and non-target attributes). 
Note that when we composed the text prompt with different attributes, we combined them with commas, in the order of object, style, and the specific attribute (if considered). 
For \textbf{directional prompt}, we calculated a directional vector between the original and additional attributes and added it to the prompt composed of the original and non-target attributes with different weights. For \textbf{prompt stencil}, we first generated an image with the prompt of the original and non-target attributes and did an overcoat with the prompt of the additional and non-target attributes. Here, we varied the level of the overcoat with the weight. When we added noise to the overcoating, we used a seed that was different from the seed we sampled before (as using the same seed resulted in a low-quality image). We fixed this seed across different overcoat weights. 
For \textbf{prompt \added{intervention}}, we first started the generation with the prompt of the original and non-target attributes. Then, at a specific moment of the generation, we changed it to the prompt of the additional and non-target attributes. In this case, with higher weights, we changed the prompt earlier. For these approaches, we used three weights, 0.25, 0.5, and 0.75, on a scale of 0 to 1. 
For \textbf{prompt concatenation}, we combined all of the original, additional, and non-target attributes in the prompt (e.g., ``the mix of cat and dog, impressionism'').
With these approaches, we generated 4368 image pairs for all target attribute types, using 50 diffusion steps with the guide scale set to 7.5.

\subsection{Metrics}

\begin{table}[t]
\caption{Options used for \texttt{addition approach} questions. Bolded text indicates the option name.}
\begin{tabular}{p{0.10\textwidth}p{0.35\textwidth}}
                                           & Options of \texttt{addition approach} questions                                                                  \\
\noalign{\global\arrayrulewidth=0.3mm}
\hline
\noalign{\global\arrayrulewidth=0.15mm}
For styles and specific visual attributes & \textsc{original attribute} and \textsc{additional attribute} are both applied in the new image, but largely to different places/things in the image. (\textbf{Separate}) \\ \hline
\multirow{7}{*}{For objects}          & \textsc{original attribute} and \textsc{additional attribute} are placed together in the new image as separate objects. (\textbf{Separate}) \\ \cline{2-2}
                                           & The new image is \textsc{original attribute}-shaped \textsc{additional attribute}. (\textbf{O-Shaped A})                                        \\\cline{2-2}
                                           & The new image is \textsc{additional attribute}-shaped \textsc{original attribute}. (\textbf{A-Shaped O})                                        \\ \hline
\multirow{7}{*}{Common}            & \textsc{additional attribute} is added to the new image, but not in the ways described above. (\textbf{Mixed})                   \\ \cline{2-2}
                                           & \textsc{additional attribute} is not added to the new image, but the image changed. (\textbf{NoMixChange})                            \\ \cline{2-2}
                                           & \textsc{additional attribute} is not added to the new image, and the image did not change. (\textbf{NoMixNoChange})                     
\end{tabular}
\Description{The first row of the table is saying "Options of addition appraoch questions." The second row has the header "for styles/specific visual attributes," with the options of "Original attribute and additional attribute are both applied in the new image, but largely to different places/things in the image (Separate)." The next three rows have the header of "For objects." They have options of "Original attribute and additional attribute are placed together in the new image as separate objects (Separate)," "The new image is Original attribute-shape Additional attribute (O-shaped A)," and "The new image is Additional attribute-shaped Original attribute (A-shaped O)." The next three rows have the header of "Common," with options of "Additional attribute is added to the new image, but not in the ways described above (Mixed)," "Additional attribute is not added to the new image, but the image changed (NoMixChange)," and "Additional attribute is not added to the new image, and the image did not change (NoMixNoChange)." }
\label{tab:addition_approach}
\end{table}

We considered four metrics: 1) how clearly the new attribute is added (\texttt{addition}), 2) how clearly the original attribute persists (\texttt{remain}), 3) how similar the newly generated image is to the original image (\texttt{similarity}), and 4) the specific way in which the new attribute is added (\texttt{addition approach}). 
For metrics other than \texttt{addition approach}, we used a 7-level Likert scale to gather answers. For the \texttt{addition approach} question, options varied depending on the types of attributes added, as in Table~\ref{tab:addition_approach}.

\subsection{Procedure}
For generated image pairs, we conducted a characterization study on Amazon Mechanical Turk. We showed each crowd worker ten pairs of randomly sampled images. 
For pairs generated with the target attribute of styles, we included examples of each style so that those who do not know the style terms can see examples. For each pair, we asked the worker four questions about all metrics. One pair in each set showed identical images and was used as an attention check. Crowd works were filtered if they answered the \texttt{similarity} metric at lower than 6 (of 7) or their answer to the \texttt{addition approach} metric was something other than NoMixNoChange (Table~\ref{tab:addition_approach}). We also filtered out a worker's answers when they answered in streaks of the same value for \texttt{addition}, \texttt{remain}, and \texttt{similarity}. Specifically, we filtered out the worker's answer if the ratio of the same value is higher than 70\%. 
We recruited workers with an acceptance rate higher than 98\% and an accepted HIT number greater than 10,000. We only recruited workers in the US and paid them \$1.50 (9-minute task, \$10/hr payment rate).

\subsection{Result}

\begin{figure}[t]
\centering
  \includegraphics[width=0.478\textwidth]{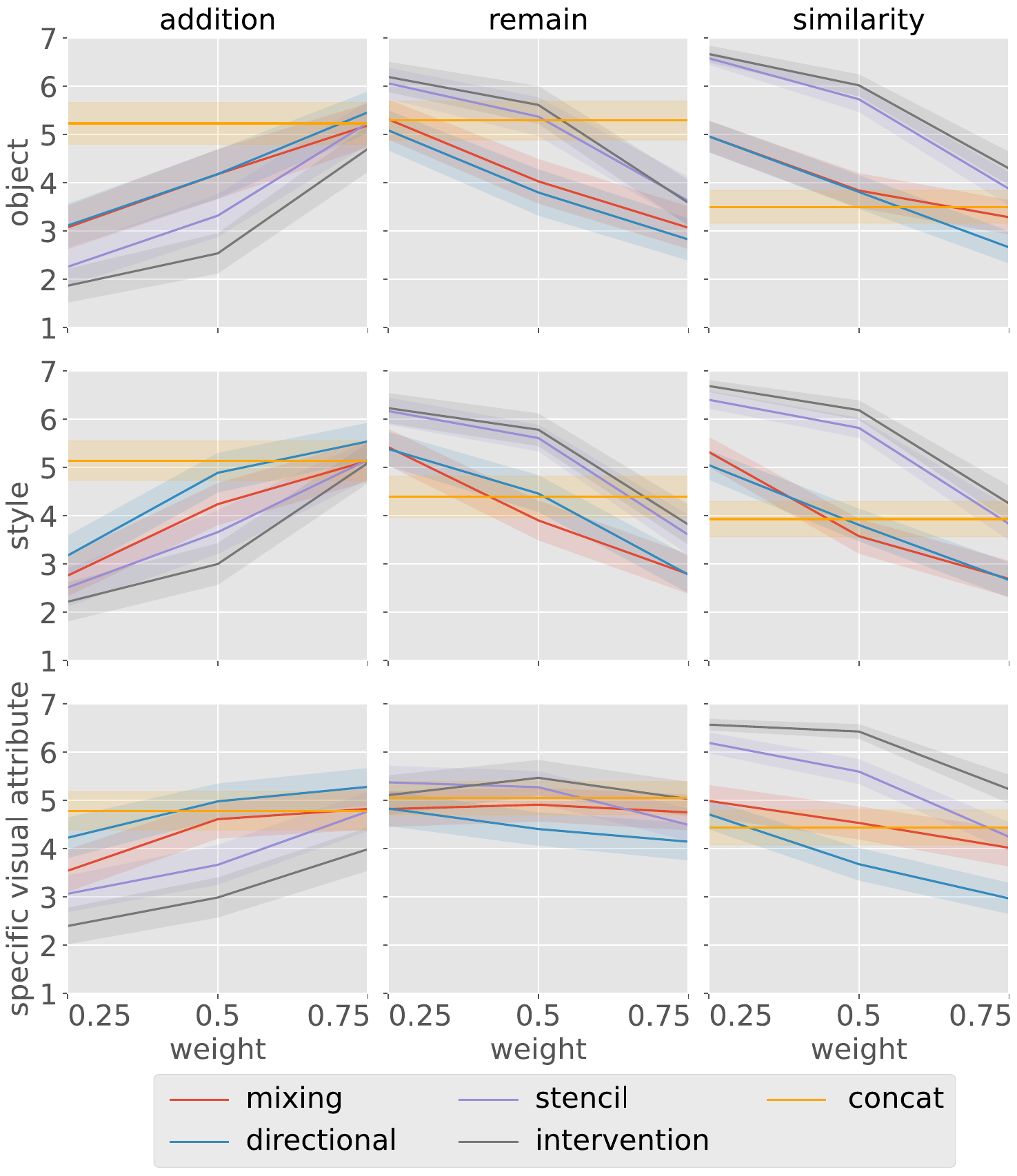}
  \caption{\texttt{addition}, \texttt{remain}, and \texttt{similarity} of different conditions and target attributes. The shaded areas indicate 95\% confidence intervals.}
  \label{fig:pp_add_remain_sim}
  \Description{There are plots in 3 by 3, which each row saying object, style, and specific visual attribute, and each column saying addition, remain, and similarity. For each plot, the x-axis stands for the weights between 0.25 to 0.75, and the y-axis stands for the score in each aspect. The plots for "addition" have the similar patterns that the score is going up with the higher with more weights. The score tends to be highest with directional prompt, followed by prompt mixing, prompt stencil, and prompt intervention. Prompt concatenation tends to have high score, but as it does not have the concept of the weight, it is horizontal. For "remain,", as the weight increases, scores tend to decrease. Score tends to be highest with prompt intervention, followed by prompt stencil, then either prompt mixing or directional prompt. Prompt concatenation tends to have a bit lower score than prompt stencil and prompt intervention, when the weight is below 0.5. It is also horizontal. For similarity, the score also tends to decrease with the increase of the weight. Prompt intervention and prompt stencil have high scores than prompt mixing and directional prompt. Moreover, prompt concatenation tends to have quite low scores for this criterium. The score is also horizontal. }
\end{figure}

\begin{figure}[t]
\centering
  \includegraphics[width=0.478\textwidth]{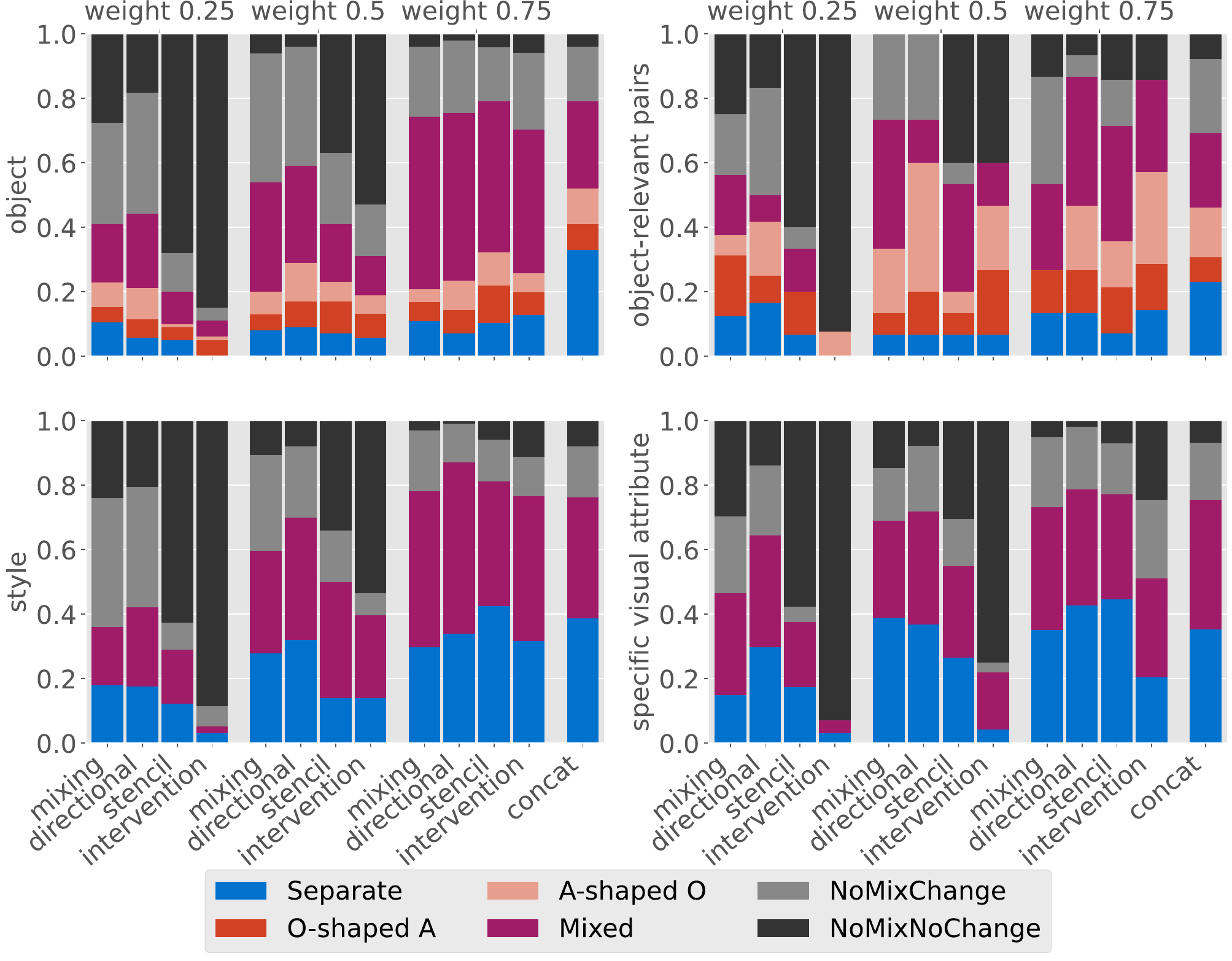}
  \caption{\texttt{addition approach} of different conditions and target attributes. For objects, we also show the results that only consider pairs consisting of closely relevant objects.}
  \label{fig:approach_rate}
  \Description{A plot showing the distribution of addition approach with different target attributes. Target attributes include object, object-relevant pairs, style, and specific visual attribute. Object and object-relevant pairs have categories of separate, O-shaped A, A-shaped O, Mixed, NoMixChange, and NoMixNoChange. Style and specific visual attribute have categories of Separate, Mixed, NoMixChange, and NoMixNoChange. For each target attribute, there are weights of 0.25, 0.5, and 0.75, with each weight having results for prompt mixing, directional prompt, prompt stencil, and prompt intervention. With low weights, the rate of categories other than NoMixChange and NoMixNoChange are higher with prompt stencil and prompt intervention than with prompt mixing and directional prompt. As the weight increases, the rate NoMixChange and NoMixNoChange becomes similar different conditions. In object-relevant pairs, the rate of O-shaped A and A-shaped O is higher than those from object. }
\end{figure}

Figure~\ref{fig:pp_add_remain_sim} shows the \texttt{addition}, \texttt{remain}, and \texttt{similarity} results for each target attribute. 
Figure~\ref{fig:approach_rate} presents how mixtures of prompts with different conditions and weights resulted in different \texttt{addition approaches}. 
For all attributes and conditions that can be weighted, the increase of weights resulted in higher \texttt{addition} while decreasing \texttt{remain} and \texttt{similarity}, with more \texttt{addition approaches} other than NoMixChange and NoMixNoChange. 
There were clear trade-offs between conditions: those approaches that more clearly add new attributes tend to lose the original attribute and the similarity to the original image. Overall, prompt \added{intervention} induced the minimal addition of the new attribute while highly maintaining the original attribute and the similarity to the original image. 
Other conditions followed in manifesting such patterns: prompt stencil, directional prompt, and prompt mixing (in that order). 
The specific trend varied between different target attributes. 
With the target attribute of objects and styles, prompt stencil and prompt \added{intervention} had similar \texttt{remain} and \texttt{similarity} scores, where prompt mixing and directional prompting formed another group. 
For specific visual attributes, \texttt{remain} and \texttt{similarity} gradually changed with weight change, which would be because the added attribute is a smaller part of the whole image. 

With \texttt{addition approaches}, prompt \added{intervention} had a low rate of changing images with low weights (i.e., high NoMixNoChange), followed by prompt stencil. For the target attributes of objects and styles, as weights increase, these rates for prompt \added{intervention} and prompt stencil increase to those of prompt mixing and directional prompt. Only for the target attribute of specific visual attributes, prompt \added{intervention} changes images in a lower rate than other conditions even with increased weights.

Concatenation could add the new attribute while remaining the original attribute, but it produced images not very similar to the original image. Moreover, concatenation of objects more frequently placed separate objects rather than mixing them. 

For objects, the distribution of \texttt{addition approaches} could be different when only considering pairs that are closely relevant (i.e., tree-river, woman-man, dog-cat, and love-hate). With relevant pairs, there were more O-shaped-A or A-shaped-O, potentially as two mixed objects were semantically relevant (sometimes, even visually). At the weight of 0.5, directional prompt and prompt \added{intervention} has high rates of O-shaped-A and A-shaped-O, while those rates for prompt stencil were low with the weight of 0.5. With high weights (0.75), prompt \added{intervention} had the highest rate of O-shaped-A and A-shaped-O.

\section{User Study}
\label{sec:user_study}
We conducted a user study to understand how \sysc{} extends the use of diffusion-based T2I models with interactions inspired by how we handle paint mediums. We focussed on how interactions of \sysc{} could affect the user's experience in exploring and steering T2I generations to ``create'' visual artifacts. Therefore, we conducted an observational study with qualitative analysis. 

\subsection{Participants}

\begin{table}[]
% \small
\centering
\caption{User study participants, with their expertise in visual arts, the domain of interest, and experience in T2I models.}
\begin{tabular}{lllll}
\hline
   & Visual art & Year & Domain                            & T2I \\ \hline
P1 & Hobbyist             & 5    & Vector arts                       & Yes  \\
P2 & Hobbyist             & 10   & Paintings, cartoons, graphic arts & No \\
P3 & Hobbyist             & 20   & Sketches, paintings               & No \\
P4 & Hobbyist             & 30   & Simple drawings, paintings        & Yes \\ 
P5 & Novice               & N/A  & N/A      & No  \\
P6 & Novice               & N/A  & N/A      & Yes  \\
P7 & Hobbyist             & 3    & Sketches & No  \\
P8 & Novice               & N/A  & N/A      & No \\
\hline
\end{tabular}
\Description{The table has description about each participant, about their level of experience in visual art, years of experience, domain of interest, and their experience in text-to-image models. The first participant, P1, is hobbyist with 5 years of experience in vector arts, and have the experience in text-to-image models. The second participant, P2, is hobbyist with 10 years of experience in paintings, carttons, and graphic arts, without experience in text-to-image models. The third participant, P3, is a hobbyist with 20 years of experience in sketches and paintings, without experiences in text-to-image models. The fourth, P4, is a hobbyist with 30 years of experience in simple drawings and paintings, with experience in text-to-image models. The fifth, P5, is a novice without experience in text-to-image models. The sixth, P6, is a novice with the experience in text-to-image models. The seventh, P7, is a hobbyist with three years of experience in sketches, without experience in text-to-image models. The eighth, P8, is a novice without experience in text-to-image models.}
\label{tab:pp_participants}
\end{table}

We recruited eight participants (five females and three males, ages 22-51, M=28, SD=9.53) through university mailing lists. We asked participants to do a prescreening survey, checking if they can participate, as the study requires participants to see and hear. 
% The prescreening survey also asked about the experience in visual arts and T2I models.  
% below a subject to change in the full study
We recruited hobbyists or novices in visual arts, as experts would be less likely to use automated generation tools in their practice (i.e., they have the expertise to create visual arts by themselves). 
% Three participants had used T2I models. 
During the study, we asked them to complete a prescreening survey that asked about their experience in visual arts and T2I models, whose results are summarized in Table~\ref{tab:pp_participants}. 
We gave each participant an Amazon gift card worth \$20.

\subsection{Procedure}

We conducted an in-person lab study. We first asked participants to complete a pre-survey. Then, we showed the participants a video with an overview of the study (5 minutes). As we asked participants to think aloud during the study, this video instructed participants about the concept and an example of think-aloud. The video also introduced the basic functions of \sysc{}, which are raster image editing functions other than image generation (e.g., brushing, erasing). Then, the video explained how to generate images with a single prompt. After the first video, we asked the participants to try the functions in \sysc{}. The participants then went through four rounds of task sessions for four functions, in the order of prompt mixing, directional prompt, prompt \added{intervention}, and prompt stencil. 
The participants went through the fixed order since the latter functions require knowledge of the previous ones. 
For each task session, participants went through four steps: 1) watching an instruction video, 2) trying out the function as a tutorial \added{with the researcher's guidance}, 3) \added{freely} creating visual artifacts as they want, while thinking aloud, and 4) completing a post-task survey. Each instruction video took 1-2 minutes. Each tutorial took about 5 minutes. 
\added{We gave 10 minutes for each creation task} and asked the participants to actively try the \added{newly learned} function.  
Post-task surveys asked participants if the function they had just tried facilitated 1) control of image generation or 2) exploration of good surprises. After all functions, participants were asked to complete an exit survey, which asks about the general usage of the tool. This survey asked questions in the creativity support index~\cite{cherry2014quantifying}, except those that questioned whether the tool helped collaboration. The post-survey also asked about the participant's sense of ownership and contribution, and if they felt they were collaborating with the system. After the post-survey, we conducted a short interview. In the interview, we asked about their strategies for using \sysc{}, how they felt about the ownership of the artifacts, and their impression of the four functions. The entire study took no more than 100 minutes. 

\subsection{Results}

We report on the results of surveys, observations of the task with think-aloud, and interviews.

\subsubsection{Survey results}
\begin{figure*}[t]
    \centering
    \includegraphics[width=\textwidth]{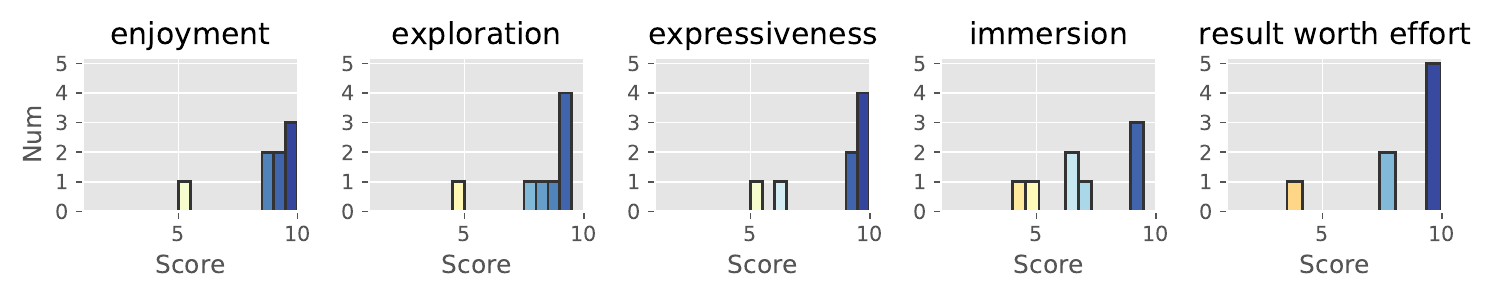}
    \caption{
        The histogram of responses on the creativity support index questions. The high score indicates that the participant perceived that \sysc{} supports the criteria. 
    }
    \Description{There are histograms for enjoyment, exploration, expressiveness, immersion, and result worth effort. Scores are mostly positive, except few cases. Enjoyment, exploration, expressiveness, and result worth effort has one user either neutral or negative, and immersion has two users either neutral or negative.}
    \label{fig:pp_csi}
\end{figure*}

Figure~\ref{fig:pp_csi} shows the results on the creativity support indexes. Participants were generally positive about \sysc{}, perceiving that it facilitated enjoyment, exploration, expressiveness, and immersion, while the results were worth their effort. However, there was one participant who responded neutrally or negatively to these questions. In this case, the participant had very concrete expectations of what they wanted. For immersion, there was one participant who answered negatively about the immersive aspect of the tool. 

\begin{figure}[t]
    \centering
    \includegraphics[width=0.478\textwidth]{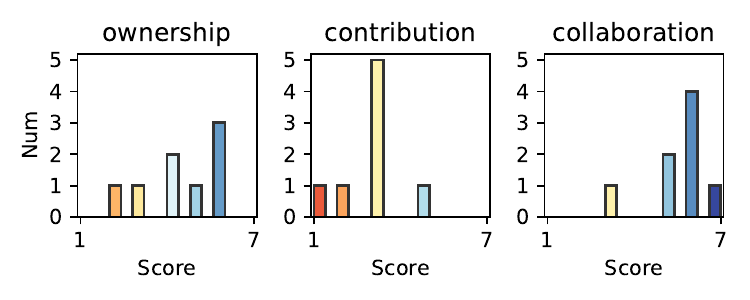}
    \caption{
        The histogram of responses to the question about the sense of ownership, contribution, and collaboration. The higher the scores, the participant felt that they have more ownership than \sysc{}, they contributed more than AI, and they collaborated with AI functions. 
    }
    \Description{Three historgrams for ownership, contribution, and collaboration are shown. Ownership is widely distributed, with a bit more people on the side of having high sense of ownership. With contribution, more people think that they contributed less. With collaboration, many people think that they collaborated with AI functions.}
    \label{fig:pp_ownership}
\end{figure}

Figure~\ref{fig:pp_ownership} shows how participants felt about ownership of the generated images, how much contribution they made (compared to AI), and whether they collaborated with \sysc{} in creating the artifact. Interestingly, participants felt that AI contributed more, but many still answered that they have some ownership of the generated artifact. At the same time, participants tended to answer that they `collaborated' with \sysc{}.

\begin{figure}[t]
    \centering
    \includegraphics[width=0.478\textwidth]{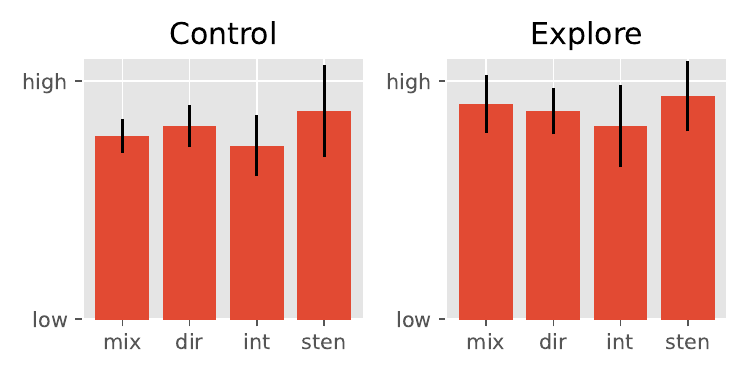}
    \caption{
        Comparison of four different functions on if they helped with 1) controlling or 2) exploring generation. \texttt{mix}, \texttt{dir}, \texttt{\added{int}}, and \texttt{sten} stand for \textit{prompt mixing}, \textit{directional prompt}, \textit{prompt \added{intervention}}, and \textit{prompt stencil}, respectively. The error bars indicate 95\% confidence intervals.
        % The red lines between conditions indicate significant differences. 
    }
    \Description{Two plots are shown for how four functions impact the user's sense of control and exploration. Participants had high sense of control and exploration for all four functions, while prompt stencil having highest score.}
    \label{fig:pp_control_explore}
\end{figure}

Figure~\ref{fig:pp_control_explore} shows the participants' perceptions on how each function supported 1) the control of generation and 2) the exploration of interesting and good surprises. Overall, participants perceived all functions positively.
While it is difficult to learn significant differences between functions due to the small size of the data, participants tend to perceive that the prompt stencil helped the most with controlling and exploring generation, while \added{intervention} prompts helped the least.

\subsubsection{Qualitative Results}
For qualitative results, we analyzed think-aloud, screen recording, and interviews by iterative coding with inductive analysis. We present findings on four functions, trade-offs in designs, the complexity of AI, and ownership issues.

\paragraph{Four functions}

\begin{figure}[t]
    \centering
    \includegraphics[width=0.478\textwidth]{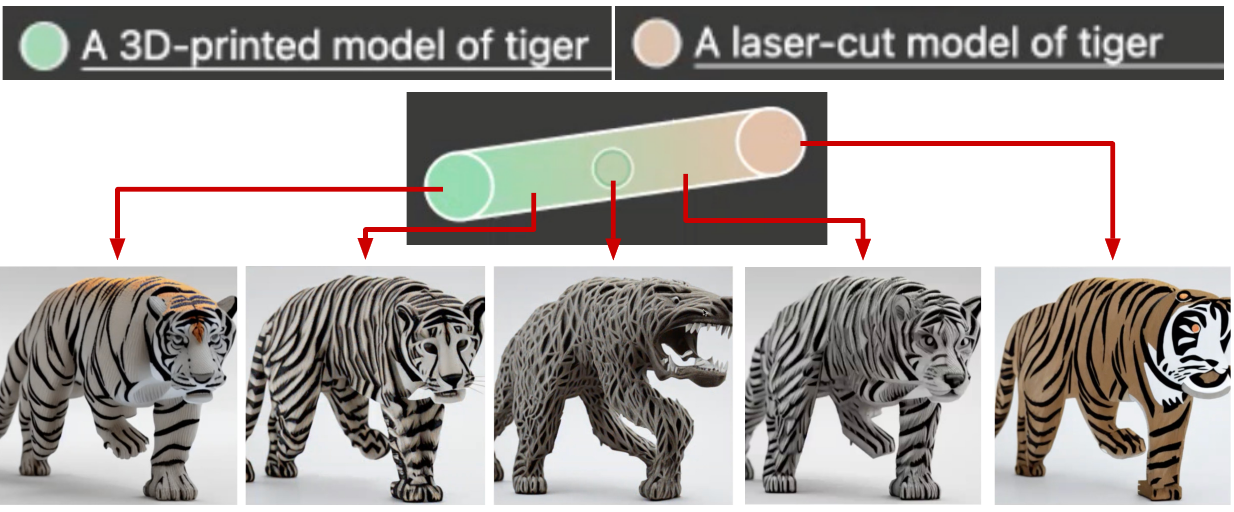}
    \caption{
        Prompt mixing from P5. 
        By mixing semantically close prompts, the user can control the generation to explore the image space in between.  
    }
    \Description{Two prompts, "A 3D-printed model of tiger" and "A laser-cut model of tiger" are mixed on the palette interface, and there are different results generated from different mix of prompts. Results with the mix of prompts tend to have both attribute from 3D-printed models and laser-cut models.}
    \label{fig:pp_interpolation_case}
\end{figure}

\begin{figure}[t]
    \centering
    \includegraphics[width=0.478\textwidth]{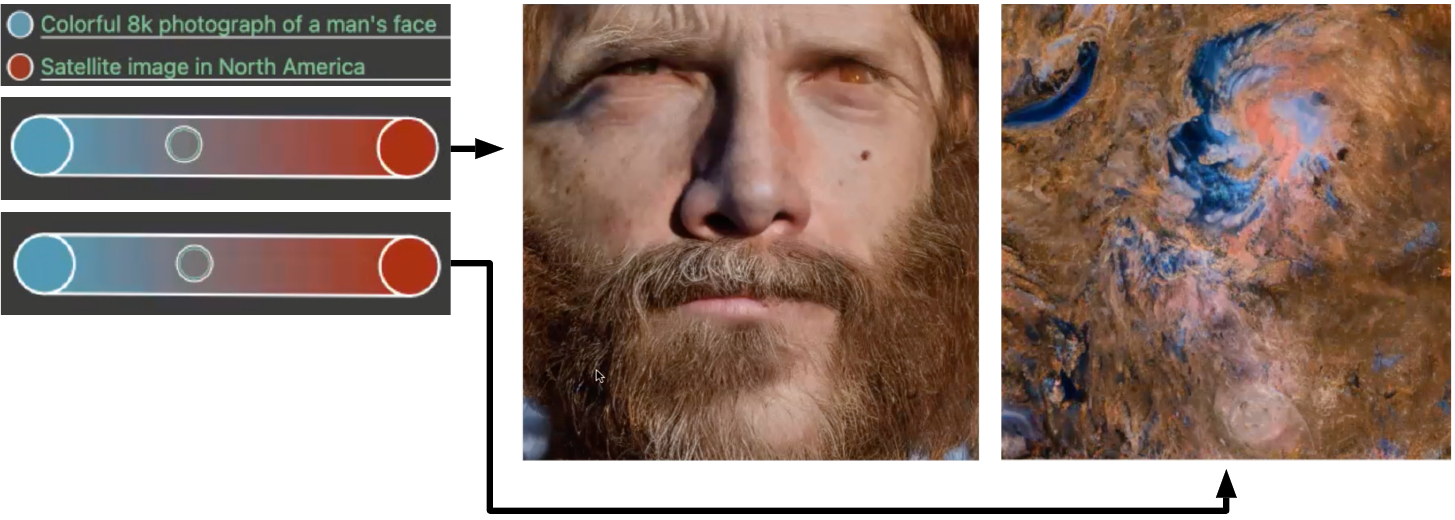}
    \caption{
        Prompt mixing from P1. 
        For semantically far prompts, interpolation of vector-embedded prompts sometimes did not result in images with a mixture of concepts. 
    }
    \Description{There are two prompts, "Colorful 8k photograph of a man's face" and "Satellite image in North America," and they are mixed on the palette interface. However, with a very change of the weight, the image drastically changes from a man's face to satellite images, without mixing thme with each other.}
    \label{fig:pp_interpolation_limitation}
\end{figure}

As seen in Figure~\ref{fig:pp_interpolation_case}, \textbf{prompt mixing} allowed participants to explore the image space that is difficult to describe verbally (N=7). 
Some participants mentioned that visualization and interactions on the Prompt Palette interface helped them explore and manipulate the prompt semantics (N=2).
One interesting thing that one of our participants (P1) discovered was that when two prompts are semantically far, mixing concepts does not change the image linearly, but more in drastic ``steps.'' For example, in Figure~\ref{fig:pp_interpolation_limitation}, P1 tried to mix two prompts, ``Colorful 8k photograph of a man's face'' and ``Satellite image in North America.'' Here, at a certain boundary, a small increase of weights on one prompt could drastically change the image, indicating that the interpolation did not impact the result linearly.
% One participant (P5) mentioned that they would like to mix more than three prompts, which was not supported in the current version of the tool. 

\begin{figure}[t]
    \centering
    \includegraphics[width=0.478\textwidth]{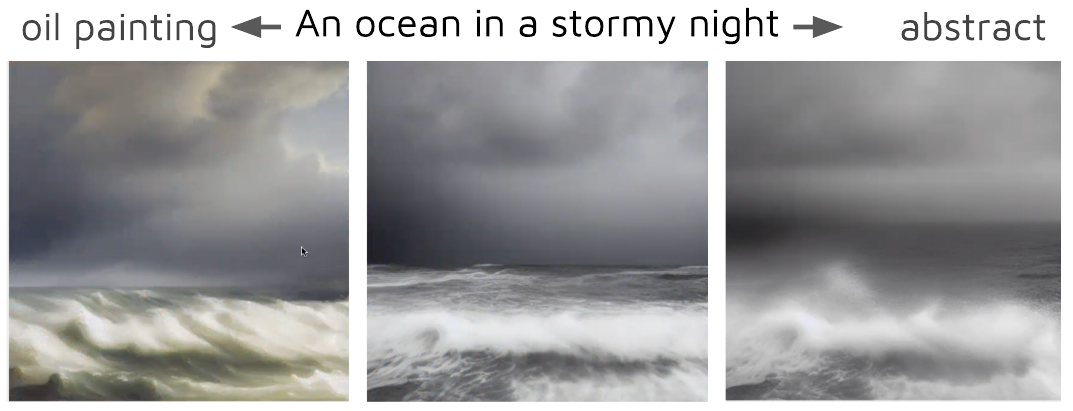}
    \caption{
        Directional prompt from P6.
    }
    \Description{There is a prompt "An ocean in a stormy night" extended with a directional prompt of "oil painting" and "abstract." The oil painting version has more brush-stroke-like textures, while abstract version has more hazy texture. }
    \label{fig:pp_direct_case_limitation}
\end{figure}

\textbf{Directional prompts} could help participants add attributes that do not exist within the first prompt they have tried (N=6, Figure~\ref{fig:pp_direct_case_limitation}). P3 mentioned that the function helped explore and do fine-grained controls between two opposite concepts: \inquote{I think the strength of this is like you can see how opposed things are, and how it is seen in between $\ldots$ Basically this function allows you to explores something that is in-between. Sometimes it's very difficult to imagine things, and this would have been very helpful in that.}
One challenge in using directional prompts was deciding on two semantically opposite prompts (N=5). The participants expected that \sysc{} could have recommended the options for opposite prompts after the user inputs one prompt. In other cases, the way the participants interpret the prompts did not align with \sysc{}'s. We return to this below.

\begin{figure*}[t]
    \centering
    \includegraphics[width=\textwidth]{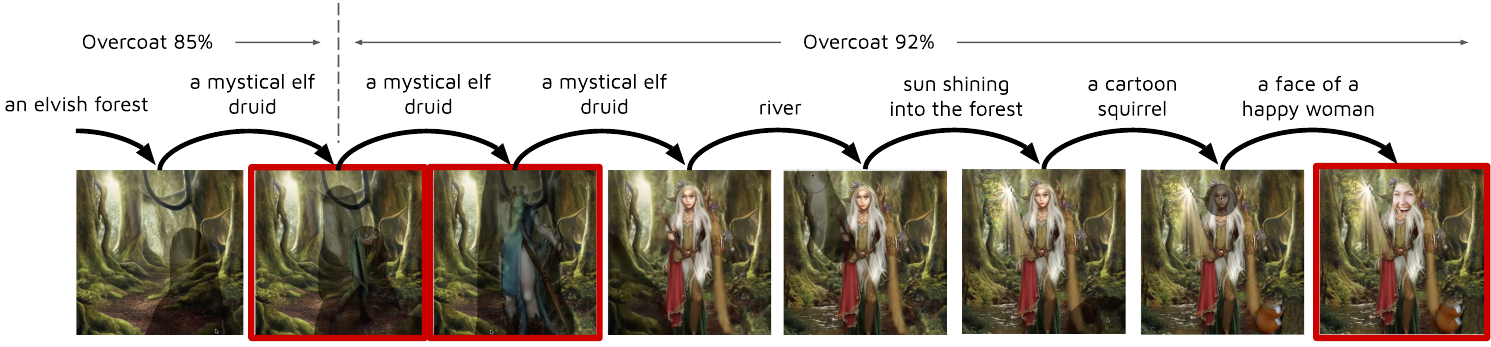}
    \caption{
        Prompt stencil from P2.
    }
    \Description{The iterative use of prompt stencil is shown. The first is "an elvish forest" on the whole canvas, and the second is "a mysical elf druid" on the right bottom, with the overcoat level of 85\%. After this point, the overcoat level of 92\% is used. As the second has failed to generate coherent druid, the user tried twice more, and finally get the image of a mystical elf druid. Then, the user "drew river", "sun shining into the forest", "a cartoon squirrel". At last, the user tried "a face of a happy woman," which failed to generate a coherent image. }
    \label{fig:pp_stencil_case}
\end{figure*}

\textbf{Prompt stencil} allowed users to perform fine-grained controls with localized image generation (N=7). As in Figure~\ref{fig:pp_stencil_case}, participants could gradually create the image by adding and changing visual elements in the scene. Participants could also adjust the overcoat level
to generate partial images that are more or less similar to the existing ones. However, the prompt stencil also had limitations. For example, as in ``a mystical elf druid'' in Figure~\ref{fig:pp_stencil_case}, newly generated parts could be incomplete. Furthermore, as in ``a face of a happy woman'' of Figure~\ref{fig:pp_stencil_case}, generated images could be mismatched with the existing ones. In some cases, the style of the newly generated images did not match the existing ones.

\begin{figure}[t]
    \centering
    \includegraphics[width=0.478\textwidth]{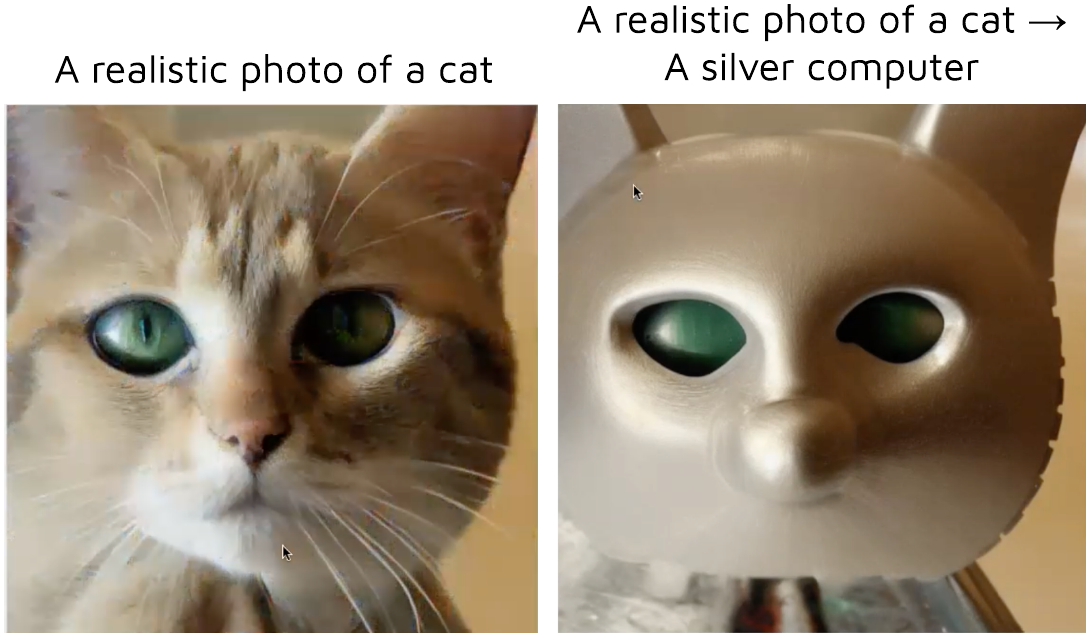}
    \caption{
        Prompt \added{intervention} from P2.
    }
    \Description{The left shows an image of "A realistic photo of a cat," and the right shows "A realistic photo of a cat" to "A silver computer," which results in a computer-textured cat.}
    \label{fig:pp_switching_case}
\end{figure}
Participants thought that \textbf{prompt \added{intervention}} allows them to generate interesting mixes of prompts (N=6). P2, who created the image in Figure~\ref{fig:pp_switching_case}, mentioned: \inquote{I think the strength of that is making something completely ridiculous and fun and changing an aspect of something to match something else.} However, prompt \added{intervention} was the most difficult to use (N=5). Some participants were unable to create a satisfactory image. It was due to the difficulty of deciding when to switch prompts as it was hard to guess the result only by seeing intermediate generation results (i.e., noisy images during the diffusion process). 
The interface showing all previous generations (in the prompt palette in Figure~\ref{fig:pp_teaser}c) could help users understand the previous generations they have tried (N=3) and iterate on the prompt \added{intervention}. However, it was easy to clutter the interface with multiple rounds of iteration.

\paragraph{Design Trade-offs}

Participants mentioned two potential trade-offs in the design of T2I generation tools. The first was the design trade-off between focusing on one canvas versus curating many results (N=2). We designed \sysc{} to allow users to iterate on a single canvas, giving users the experience closer to ``gradually creating an image.'' However, due to stochasticity in the diffusion model (e.g., randomness from different seeds), participants found that seeing multiple results would be helpful in some cases. 
Furthermore, we designed \sysc{} to allow users to have more controls and interventions. For example, the tool allows users to have a high degree of freedom to change the prompts during generation. Although such designs open up new interactions in using diffusion-based T2I tools, some users found such designs too manual (N=2). Participants mentioned that the balance between automation and manual interventions would help.

\paragraph{High Complexity and Randomness of AI}
\label{sec:pp_high_comp_rand}

Participants mentioned that the high complexity and randomness of AI behaviors were limitations of \sysc{} (N=7). 
Such complexity and randomness could make the generation result misaligned from the user's intention. Dissatisfaction tends to be more intense when the user has a more concrete picture of what they want. For example, P7 thought that prompt stencils often failed to generate image parts with consistent perspectives and was most dissatisfied with the prompt stencil.
To facilitate generative AI even with such barriers, participants adopted some strategies. Some tried to understand how AI works in simple settings (e.g., using a single prompt) and then applied more complex functions (e.g., prompt mixing) based on their understanding (N=2). Some tried to understand how the model ``interpret'' prompts by using functions that interpolate or \added{shift} the semantics of the prompts (N=2). For example, P1 interpolated the prompts of an apple and a pear to learn how the machine interprets the attributes of each fruit. 

\paragraph{Ownership and Contributions}

Participants felt some ownership of the resulting visual images (N=8), with varying degrees between participants. They mentioned that they contributed high-level ideas, while AI contributed low-level ideas and implementations. 
P1 mentioned that they became like ``Steve Jobs'' and AI would be ``an Apple employee,'' and P2 thought that using \sysc{} felt like doing an art commission with more control. 
P5 felt less ownership of the resulting piece because they were a novice in visual arts. P5 mentioned: \inquote{I think AI contributes more than me and it's because I'm a novice. I did not paint at all, and I don't use any drawing software as well. So I think all the beautiful images are created by AI instead of me. I just specify the position, and it's just parameters.}
Some participants also mentioned that they felt more ownership in the resulting artifacts if they align with what they expected (N=2).
One participant mentioned the potential issues in the legal ownership of the generated artifacts, showing concerns about the copyright. 
\section{Discussion}
Based on the suggested design of generative tools and \sysc{}, we discuss 1) the generalizability of paint-medium-like interactions in generative tools, 2) the characterization of different approaches to mixing prompts, 3) in-generation interactions for T2I models, 4) design trade-offs in generative tools, 5) ownership issues, \added{and 6) limitations}. 

\subsection{Paint-like Interactions for Generative Tools}

We can apply the idea of paint-medium-like interactions beyond \sysc{}. 
For mixing discrete semantics, we can easily replace prompts with other inputs, such as examples. 
On the other hand, we would need to redesign modularized generation specifications for each content modality. 
For example, for the generation of 3D models~\cite{jain2021dreamfields, poole2022dreamfusion}, users would need to be able to select 3D parts to iterate. 
Similarly, for content with sequential axes, such as text, video, or music, the generation specification would need to consider the sequential dimension~\cite{chung2021talebrush}. 
In the interface, they can be instantiated in sketches of different semantics along the sequential axis.
Interactions for modularized specifications would likely be more complex for mediums with both spatial and sequential dimensions (e.g., videos). Still, the design pattern of applying different semantics (analogically, colors) to different parts of the artifact would generally hold across modalities. 

For in-generation interventions, in \sysc{}, diffusion-based T2I models used the earlier prompts to decide the overall form and colors while using the later ones to render details. Similarly to diffusion-based T2I models, models for other mediums can be designed to gradually generate from ``high-level characteristics'' to ``details'' to allow user interventions in a generation. For example, music generation algorithms can generate, in the order of song structures, bars with chords, notes in each bar, and then embellishments such as legato or staccato.

\subsection{Characterizing Approaches to Mix Prompts}
Our characterization study revealed the pros and cons of approaches to mix prompts. Prompt \added{intervention} and prompt stencil tend to maintain the original attribute and similarity to the original image, while prompt mixing and directional prompt tend to add the new attribute with higher chances. All these approaches also have benefits over concatenating prompts, as the user can adjust how much of the new attribute to add. This characterization would guide us in deciding the mixing approach that would best achieve a user's specific purpose. 
We argue that researchers need to conduct this type of characterization for emerging T2I techniques, as with many different approaches, we do not yet have a good understanding of which would best fulfill a user's specific intention.

\subsection{Interaction for T2I models}

\sysc{} allows users to interact with generative models \textit{during the generation process} by changing the prompts, with earlier prompts forming the overall composition and later prompts deciding on details. While participants found this function interesting, they struggled to learn the best way to use it (specifically, for finding the right moment to change the prompt). 
Seeing and interpreting intermediate representations might help overcome the limitations. For example, if the intermediate noise-added image has quite a concrete object, it might indicate that changing the prompt would not induce changes. 
However, not many users are familiar with such noise-added images. 
Therefore, users would need to \textit{learn} to interpret noisy images, which places more load on users. 
Making intermediate results more understandable to human users would be an approach to facilitate in-generation interactions for diffusion-based T2I models~\cite{bansal2022cold}. For example, diffusion models that gradually concretize images from more pixelated ones would be more understandable to the users, allowing them to grasp what the model might generate from the current intermediate step. Moreover, such representations can allow users to edit the intermediate results. For example, with pixelated intermediate images, if the user is generating a human face and spots a ``blonde'' color in the area of hair, they would be able to change the color of the hair by changing the region to other colors.

We also emphasize that T2I models are quickly evolving, and \sysc{} can be extended to new models.
\added{For example, with models designed specifically for overcoating and inpainting~\cite{suvorov2021resolution}, prompt stencil interaction can be improved, and the user reaction might likely change.} 
In \sysc{}, we did not include negative prompts~\cite{negativeprompt2023}. This can be adopted into our interface, as either a text box that applies to all of the user's generation or as another palette that allows users to flexibly define the negative semantics. 
\sysc{} also does not include approaches that add structural conditions, such as ControlNet~\cite{zhang2023adding}. Again, this also can be included either by allowing the specification of those conditions during prompt stencil or by training a ControlNet model that can condition the generation with the rough stencils. 

\subsection{Design Trade-Offs in Generative Tools}
% curation + iteration
% manual vs automation
From the user study, we found design trade-offs for generative tools. First, while ``creation tools'' often assume a single artifact to be created (e.g., a single canvas for image editors), due to the complexity and randomness in generative models, generative tools would require some ``curation'' of multiple results. Providing both features would allow steering experiences while addressing some issues with the randomness of algorithms\added{~\cite{koyama2020sequential, chen2011nonlinear}}. For example, a generative tool can have multiple rounds of interactions that first receive user specifications, generate a set of candidates, allow users to select one of them, and then iterate. 
For effective steering, other control approaches, such as giving structural information of image renditions~\cite{zhang2023adding}, can be adopted.
For effective curation, it would be valuable to learn the user's preferences during the interactions to better align curated results with the user preferences. 
The second trade-off is the balance between automation and manual controls. With this trade-off, simple and automated interactions can be a ``low threshold'' way to steer the generation, while more manual steering interactions can be a ``high ceiling'' option~\cite{resnick2005design}. 

% ownership
\subsection{Ownership of Generated Artifacts}

Users of \sysc{} had some sense of ownership of the artifacts generated, as they contributed high-level ideas. At the same time, as \sysc{} contributed ideas and implementations of lower levels, they would have felt less ownership than creating artifacts themselves. For generative creation tools to secure the user's sense of ownership regarding the final artifacts, it would be important to understand which aspects of artifact creation contribute to the sense of ownership. For users who do not put a lot of value on manual labor, automating some parts of the artifact creation would not hurt the user's sense of ownership much. However, if the user values the skills and efforts involved in the creation of artifacts, then automation would hurt the sense of ownership. Hence, the tool would need to understand the user's values and allow users to select for which part they want to use generative AI. Ultimately, we need to incorporate generative functions into existing workflows so that we can preserve user values in their workflow~\cite{yan2022flatmagic}. 

Ownership is also a legal issue, such as not hurting copyrights. As existing diffusion models could copy content from the training dataset~\cite{somepalli2022diffusion}, it would be crucial to carefully curate the training dataset so that users do not infringe the legal ownership during their use. Although researchers have started to exclude images if the original owners do not want their images in the dataset\footnote{https://haveibeentrained.com/}, it is still opt-out. Moreover, there could be some trade-offs between preserving legal ownership and having a large-scale dataset. 
Potentially, the transformation of image data can be a way to balance the preservation of ownership and the scale of the data.

\subsection{Limitations}

\added{We did not compare interactions of \sysc{} to those of other existing tools, as our studies focused on 1) comparing different functions in combining semantics in prompts (Section~\ref{sec:characterization}) and 2) qualitatively studying usage patterns with paint medium-like interactions (Section~\ref{sec:user_study}). As many relevant T2I tools keep arising~\cite{automatic1111_2023, adobefirefly, dreamstudio}, doing systematic analysis on different interaction modes would be necessary for future work. Our qualitative study results can inform specific future study designs. For example, users perceived the value of iterative interactions in our tool while acknowledging the benefit of seeing many results at once. Based on this, future comparative studies can consider two dimensions, one being ``facilitation of iteration'' and the other being ``showing multiple results.''}
\section{Conclusion}

In this paper, we introduce an approach for interacting with generative models as if prompts were paint colors.
This design approach allows users to explore the semantic vector space in a way similar to how we mix colors. They can also gradually build the artifact with different semantics in a way similar to how we apply colors to varying parts of the painting process and the canvas. 
We are motivated by a desire to make end-to-end use of generation models more flexible and gradual with iterative steering. We apply the design approach in diffusion-based T2I models and introduce \sysc{}. \sysc{} adopts four steering approaches, prompt mixing, directional prompts, prompt \added{intervention}, and prompt stencil. Through user studies, we characterize these approaches and identify how people use the suggested interactions. Based on the findings, we draw insights into how we should design and build future generative tools.
\begin{acks}
We thank Nikola Banovic and Anhong Guo for their valuable feedback on this work.
\end{acks}

\bibliographystyle{ACM-Reference-Format}
\bibliography{main}
\balance
\appendix

\end{document}